*Recent Progress on Multiferroic Hexagonal Rare-Earth Ferrites (h-RFeO$_3$, R = Y, Dy-Lu)*

*Xin Li,[1] Yu Yun,[1,2]\* Xiaoshan Xu[1,3]\**

[1]Department of Physics and Astronomy, University of Nebraska, Lincoln, Nebraska 68588, USA
[2]Department of Mechanical Engineering & Mechanics, Drexel University, Philadelphia, PA 19104-2875, USA
[3]Nebraska Center for Materials and Nanoscience, University of Nebraska, Lincoln, Nebraska 68588, USA

\*Corresponding authors: yy549@drexel.edu (YuYun); xiaoshan.xu@unl.edu (Xiaoshan Xu)

**Abstract:**

Multiferroic hexagonal rare-earth ferrites (h-$R$FeO$_3$, $R$=Sc, Y, and rare earth), in which the improper ferroelectricity and canted antiferromagnetism coexist, have been advocated as promising candidates to pursue the room-temperature multiferroics, because of strong spin-spin interaction. The strong interactions between the ferroic orders and the structural distortions are appealing for high-density, energy-efficient electronic devices. Over the past decade, remarkable advances in atomic-scale synthesis, characterization, and material modeling enable the significant progresses in the understanding and manipulation of ferroic orders and their couplings in h-$R$FeO$_3$ thin films. These results reveal a physical picture of rich ferroelectric and magnetic phenomena interconnected by a set of structural distortions and spin-lattice couplings, which provides guidance for the control of ferroic orders down to the nano scale and the discovery of novel physical phenomena. This review focus on state-of-the-art studies in complex phenomena related to the ferroelectricity and magnetism as well as the magnetoelectric couplings in multiferroic h-$R$FeO$_3$, based on mostly the recent experimental efforts, aiming to stimulate fresh ideas in this field.

## 1. Introduction and basics

With the developments of Big Data and Internet of Things, there has been a widespread and growing demand for a device paradigm shift toward high-integration, energy-efficient, and multifunctional devices. Among the pepped up advanced materials, multiferroic materials, based on spins, electric dipoles, and elastic distortions, have attracted tremendous attention due to their exotic physical properties and novel functionalities on the macroscopic and microscopic level, holding great promise for future low-power electronic products.[1-5] The coexistence of magnetism and ferroelectricity in a single-phase material, however, is rare in nature, especially above room temperature [6-10]. Over the past decades, significant efforts have been devoted on multiferroic complex oxides to understand the origin of the ferroic orders and their couplings toward realizing strong magnetoelectric couplings at high temperatures [11-14]. A plethora of phenomena have been discovered on different levels, such as polar vortex[15-17], domain-wall conductivity[18-21], magnetoelectric coupling[22], which have expanded the conventional understanding of ferroic orders and driven the exploration of novel physical phenomena and functionalities which may be harnessed for practical multifunctional device applications [23-25].

Generally, single-phase multiferroic compounds can be classified as two types, type-I and type-II multiferroics, depending on whether the ferroelectric and magnetic states appear independently.[26] In particular, type-II multiferroics refer in particular to magnetism-driven multiferroics, wherein the magnetic order leads to an inversion symmetry breaking and thus generates the ferroelectricity. All other multiferroics can be categorized as type-I multiferroics, in which the ferroelectricity usually is stabilized at higher temperature, while the appearance of magnetic order is observed at lower temperature. Hexagonal rare-earth ferrites (h-$R$FeO$_3$, $R$=Sc, Y, and rare earth), which are categorized as type-I multiferroics, have the isomorphic crystal structure with that of h-$R$MnO$_3$ [27-29]. In comparison with their manganites counterparts, h-$R$FeO$_3$ have stronger magnetic order due to the enhanced exchange interaction between Fe sites and exhibit finite spontaneous magnetizations [30-32]. Among the most intriguing is that both spontaneous polarization and magnetization originate from the same non-polar structural distortion. Consequently, the structural distortion has been explored as a medium to couple magnetism and ferroelectricity [33], promising a highly designable and tunable multiferroicity. For instance, multiple means have been used to tune the structural distortion in epitaxial thin films, such as interface, strain, and doping [34-39].

The hexagonal rare-earth ferrites exhibit distinct crystal structure and symmetry from their perovskite counterparts. The hexagonal crystal structure (space group $P6_3cm$) serves as the foundation for the polar and spin orders in multiferroic h-$R$FeO$_3$. The unit cell of h-$R$FeO$_3$ is composed of the triangular lattice of FeO$_5$ bipyramids, which is sandwiched by the rare earth layers, as shown in Fig. 1(a). For two consecutive FeO$_5$ layers, the two triangle lattices rotate 60° with respect to each other. The lattice distortion relative to the high-temperature paraelectric phase structure (space group $P6_3/mmc$) can be decomposed into three modes [40, 41]. The $K_3$ mode corresponds to rotation of FeO$_5$ bipyramids, which can be quantified by the magnitude ($Q$) and rotation angle ($\phi_Q$) of apical oxygen (see Fig. 1(a)), associated with the corrugation of rare earth layer (see Fig. 1(b)); $\Gamma_2^-$ mode corresponds to the displacement of atoms along the $c$ axis that generates a dipole moment; $K_1$ mode corresponds to the in-plane displacement of Fe in the FeO$_5$ bipyramids, reducing their point-group symmetry from D$_{3h}$ to C$_S$ [30-32].

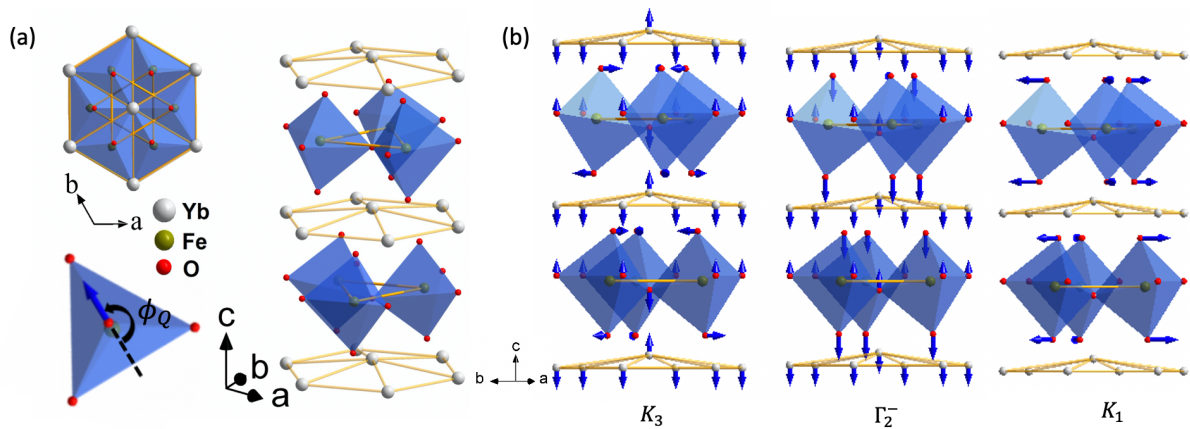

**Fig. 1** (a) Schematic of atomic structure of h-$R$FeO$_3$. (b) The structure distortions for $K_3$, $\Gamma_2^-$, and $K_1$ mode, respectively.

The non-polar $K_3$ mode induces the polar $\Gamma_2^-$ mode which leads to a buckling of oxygen polyhedral (FeO$_5$ bipyramid) and trimerization of apical oxygen anions, consequently giving rise to an offset of rare-earth cations and resulting spontaneous polarization. Thus, the spontaneous polarization is proportional to the magnitude of the $\Gamma_2^-$ mode below Curie temperature ($T_c$) of ~1000 K [42,43]. So, the ferroelectricity of h-$R$FeO$_3$ is called 'improper' or geometric ferroelectricity [44]. On the other hand, for magnetism, the $K_3$ mode breaks the three-fold rotational symmetry of the interlayer exchange interaction and determines the three-dimensional magnetic ordering temperature [30-32]. The in-plane spins along the (100) direction is energetically more favorable than the out-of-plane spins owing to their interlayer exchange interaction and $K_1$ distortion[45]. Rotation of the FeO$_5$ bipyramids leads to the canting of the Fe spins toward the out-of-plane direction due to the spin-orbit coupling and Dzyaloshinskii–Moriya (DM) interactions [33], corresponding to a weak ferromagnetism or canted antiferromagnetism. Therefore, the $K_3$ distortion, as a lattice degree of freedom, bridges the ferroelectricity and magnetism in h-$R$FeO$_3$.

Despite great promises and fascinating features, unlike their manganites cousins, the ground state of $R$FeO$_3$ ferrites in bulk are commonly orthorhombic phase rather than hexagonal phase, which is metastable state[46]. This hinders the understanding of physical mechanism and the discovery and control of new functionalities in h-$R$FeO$_3$. Strain engineering or mechanical boundary conditions, which is imposed by the semi-infinite substrate, such as YSZ (111) and Al$_2$O$_3$ (0001) substrates [30, 47], is leveraged to stabilize the metastable hexagonal phase of $R$FeO$_3$, providing a pathway to study the fundamental physics mainly about magnetism and crystal structure. More recently, advances in atomic-scale thin film growth, structural design, and characterization technique enable a electrical measurements with an oxide bottom electrode and facilitate a large degree of control over ferroelectric function, offering new opportunities to further understanding and engineering their ferroelectric, magnetic, and multiferroic properties in concert with multiscale modeling. Although significant efforts are being devoted, a comprehensive overview is still missing on recent developments and breakthroughs in hexagonal rare-earth ferrites.

In this review, we focus on the advancement of understanding on the physical mechanisms of multiferroicity from the viewpoint of experimentalist, in contrast to previous reviews which focus more on stabilization of the hexagonal structure of $R$FeO$_3$ and characterization of basic ferroic properties [46,48,49]. More explicitly, we will discuss experimental advances that lead to more insights on the underpinning mechanisms of improper ferroelectricity in ultrathin films, ferroelectric switching dynamics, as well as on the importance of spin-lattice coupling for almost all aspects of magnetism, including magnetic ordering temperature, single-ion magnetic anisotropy, and weak-ferromagnetism.

## 2. Improper ferroelectricity

The improper ferroelectricity of h-$R$FeO$_3$ is expected to be similar to that in h-$R$MnO$_3$, which has been established decades ago [29,50]. The implication of improper ferroelectricity, including its effect on depolarization field and the consequential absence of critical thickness, as well as its behavior under strain and at interfaces, are intriguing. The recent advances in atomic-scale modulation of h-$R$FeO$_3$ in epitaxial thin films, as well as the combination of phenomenological models and experiments, push forward the understanding on these implications, highlighting the application potentials of h-$R$FeO$_3$ as improper ferroelectrics toward device miniaturization and high energy efficiency.

### 2.1 Landau theory for improper ferroelectricity in h-$R$FeO$_3$

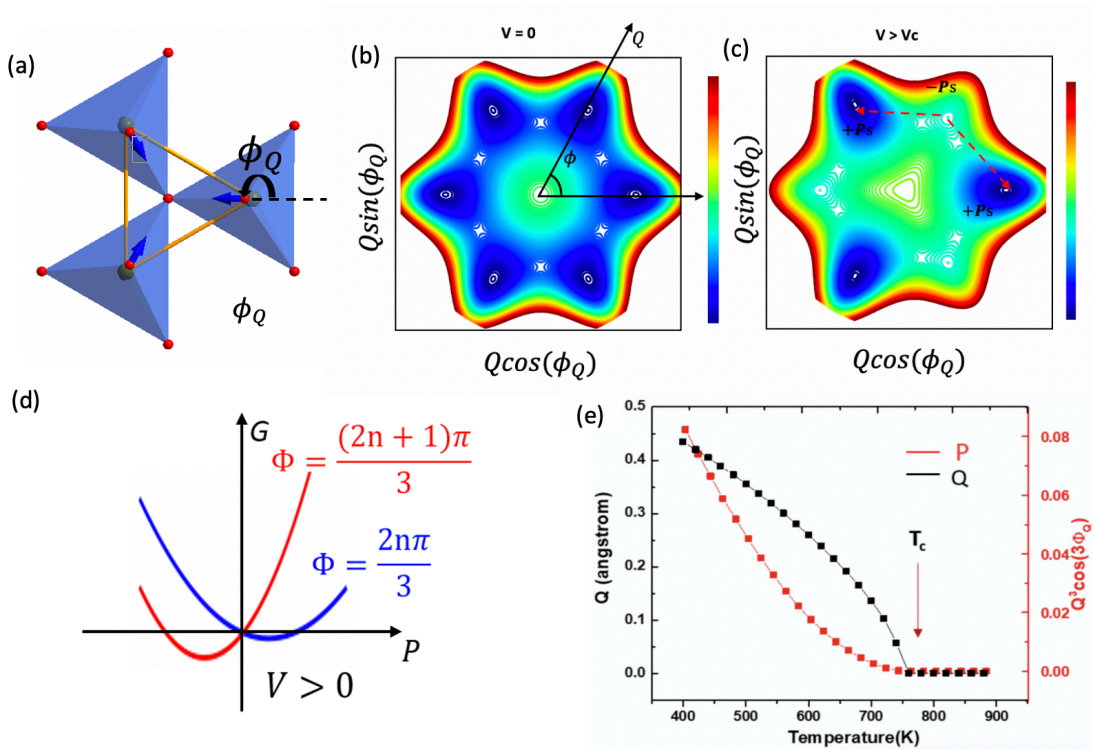

**Fig. 2** (a) The definition of order parameters of K$_3$ distortion. (b) The energy landscape of *h-R*FeO$_3$ without and (c) with electric field. (d) Schematic polarization-dependent free energy with constant phase angle. (e) The temperature-dependent paraelectric-ferroelectric phase transition for h-*Sc*FeO$_3$. (e) Reprinted with permission from ref 53. Copyright 2023. American Physical Society.

The order parameter of the improper ferroelectricity of h-*R*FeO$_3$ is the nonpolar K$_3$ distortion, in which the collective rotation of the FeO$_5$ bipyramids contributes to the geometric field to the nearby rare earth layer [50], leading to the spontaneous polarization (polar $\Gamma_2^-$ mode). Fig. 2(a) shows the part of K$_3$ distortion in terms of the displacement of apical oxygen in FeO$_5$; the direction of the displacement is defined by the angle $\phi_Q$. The free energy from Landau theory has been constructed as [51]:

$$G = \frac{a}{2}Q^2 + \frac{b}{4}Q^4 - gQ^3 P \cos(3\phi_Q) + \frac{g'}{2}Q^2 P^2 + \frac{a_p}{2}P^2 - EP \qquad (1)$$

where *Q* is magnitude of the K$_3$ distortion, *P* is the polarization, *a*, *b*, *g*, *g'*, *a$_p$* are coefficients. The first and second terms correspond to K$_3$ mode only, the third and fourth terms describe the coupling between the K$_3$ and the $\Gamma_2^-$ mode, the fifth term is for the $\Gamma_2^-$ mode, and the last term corresponds to electrostatic energy. Without electric field, the energy landscape has six-fold symmetry with energy minimum at $\phi_Q = n\frac{\pi}{3}$ (n is integer), as shown in Fig. 2(b). The polarization can be found as:

$$P = \frac{gQ^3\cos(3\phi_Q)}{g'Q^2+a_P} \qquad (2)$$

With electric field, the symmetry of the energy landscape reduces to a three-fold rotational symmetry (see Fig. 2(c)) [52]. Moreover, compared to the polarization-dependent double-well free energy landscape in proper ferroelectrics, the free energy of h-$R$FeO$_3$ depends on polarization, $Q$, and $\phi_Q$ (see Fig. 2(d)), since the primary order parameter is the structural distortion rather than electrical polarization. Nevertheless, the transition from the ferroelectric phase to the paraelectric phase still corresponds to the vanishing order parameter[40]. In other words, the Landau coefficient in first term can be approximately written as $a(T_c - T)$. Above $T_c$, the energy minimum is at $Q = 0$, corresponding to the paraelectric phase with space group $P6_3/mmc$. Fig. 2(e) shows the temperature-dependent $Q$ and $P$ from the Landau theory [53].

## 2.2 Impacts of depolarization field on improper ferroelectricity in h-$R$FeO$_3$ films

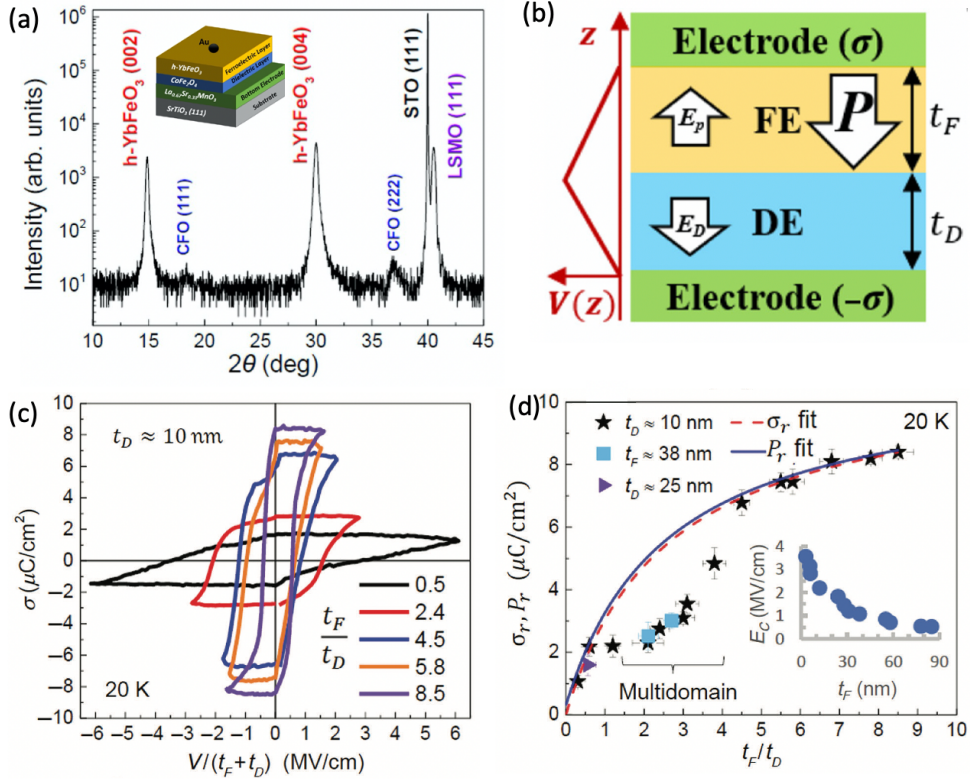

**Fig. 3** (a) X-ray $\theta$-$2\theta$ scan for h-YbFeO$_3$/ CoFe$_2$O$_4$/LSMO/STO (111) film. (b) Schematic diagram for the ferroelectric / dielectric system and related potential distribution. (c) Ferroelectric hysteresis loops of h-YbFeO$_3$ with various ratios between the thickness of ferroelectric and dielectric layers at 20 K. (d) The thickness-dependent remanent polarization and related fit for h-YbFeO$_3$ films, the inset is thickness-dependent coercive field. Reprinted with permission from ref.59. Copyright 2022. American Physical Society.

The improper ferroelectricity in h-$R$FeO$_3$ implies the insensitivity of order parameter (K$_3$ distortion) to the depolarization field and the potential absence of critical thickness [54,55], since its primary order parameter is structural distortion, in contrast to proper ferroelectrics [56-58]. Experimentally, the depolarization field can be manipulated by inserting the dielectric (DE) layer between bottom electrode and ferroelectric (FE) layer. As shown in Fig. 3(a), the epitaxial film of h-YbFeO$_3$ was grown on CoFe$_2$O$_4$ (111)/ La$_{2/3}$Sr$_{1/3}$MnO$_3$(LSMO)/SrTiO$_3$(STO)(111), in which the CoFe$_2$O$_4$ layer serves as both the dielectric layer to tune depolarization field and the buffer layer to mitigate the lattice mismatch with LSMO [58]. Fig. 2(b) shows the schematic of the potential distribution in a FE/DE bilayer system under the short-circuit condition. The ferroelectric hysteresis at cryogenic temperature (20 K), as shown in Fig. 3(c), indicates that the lower FE/DE thickness ratio ($t_F/t_D$) would suppress the remnant polarization ($P_r$) and enhance the coercive field, which is qualitatively similar to the trend in proper ferroelectrics. Moreover, the suppression of $P_r$ with decreasing $t_F/t_D$ does not follows the model for single-domain case directly, and  the deviation of experimental results from theoretical fits suggests that the h-YbFeO$_3$ films enter the multidomain states when $t_F/t_D$ is between 1 and 5 (see Fig. 3 (d)). This manifests that domain structures of improper ferroelectrics are also influenced by electrical boundary conditions. In addition, when the CoFe$_2$O$_4$ is fixed at ~10 nm, the polarization for the thinnest h-YbFeO$_3$ ($\approx$ 3 u.c.) films in Fig. 3(d) is still finite, suggesting absence of critical thickness.

## 2.3 Interfacial clamping in ultrathin h-$R$FeO$_3$ films

Although h-$R$FeO$_3$ is believed to have no critical thickness [54], the interfacial clamping effect causes a "practical" critical thickness below which the order parameter $Q$ of the K$_3$ distortion vanishes[55,59]. Fig. 4(a) shows the schematic phase diagram for the phase transition with temperature and thickness as two independent variables. Fig.4 (b) and (c) demonstrate the cross-section HADDF-STEM images for h-LuFeO$_3$/YSZ[60] and h-YbFeO$_3$/CFO interfaces [59], in which the corrugation of initial rare earth layer is greatly suppressed; the thickness dependent $Q'$ (the displacement of the $R$ atoms which is proportional to $Q$) calculated from the image of Fig. 4(c) is presented in Fig. 4(d). Thus, beyond electrical boundary condition, mechanical boundary condition also has a remarkable impact on the structure and electrical property of improper ferroelectrics. The suppression of $Q'$ near the interface by interfacial clamping could induce additional elastic energy, which can be expressed as [53]:

$$f_{elastic} = k \left(\frac{\partial Q}{\partial z}\right)^2 \qquad (3)$$

in which $k$ is the stiffness coefficient, $z$ is along $c$-axis of h-RFeO$_3$. Combining with free energy in eqt. (1), the thickness-dependent $Q$ follows the equation:

$$Q(z,T) = Q_\infty(T) \frac{1-\exp\left(-\frac{z}{\zeta(T)}\right)}{1+\exp\left(-\frac{z}{\zeta(T)}\right)}, \text{with} \begin{cases} Q_\infty(T) = \sqrt{\frac{-a_0}{b}}\sqrt{1-\frac{T}{T_s}} = Q_0\sqrt{1-\frac{T}{T_s}} \\ \zeta(T) = \sqrt{\frac{k}{-a_0}}\sqrt{\frac{T_s}{T_s-T}} = \zeta_0\sqrt{\frac{T_s}{T_s-T}} \end{cases} \quad (4)$$

in which $z$ is the film thickness, $T$ is the temperature. $a_0<0$ and $b>0$ are coefficients of Landau theory, $k$ is the stiffness coefficient at the interface, $T_s$ is the Curie temperature of bulk state, and $\zeta_0$ is the characteristic length of interfacial clamping at $T = 0$ K. The thickness and temperature-dependent phase transition can be further described as:

$$\alpha(t_F, T) = T - T_S\left[1 - 2\zeta_0^2 \frac{C_2(t_F,T)}{C_1(t_F,T)}\right] = 0 \quad (5)$$

with

$$\begin{cases} C_1(t_F,T) = \int_0^{t_F} \left(\frac{1-\exp\left(-\frac{z}{\zeta(T)}\right)}{1+\exp\left(-\frac{z}{\zeta(T)}\right)}\right)^2 dz \\ C_2(t_F,T) = \int_0^{t_F} \frac{\left(\frac{2}{\zeta}*\exp\left(-\frac{z}{\zeta(T)}\right)\right)^2}{\left(1+\exp\left(-\frac{z}{\zeta(T)}\right)\right)^4} dz \end{cases} \quad (6)$$

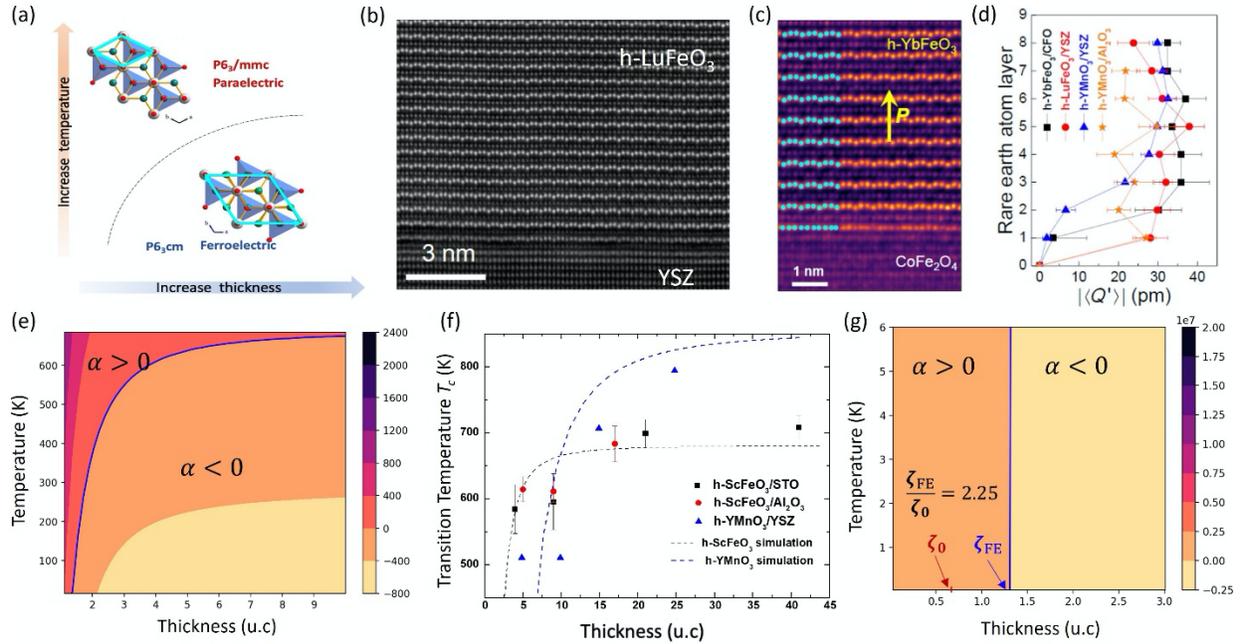

**Fig. 4** (a) Schematic for the temperature and thickness-dependent paraelectric-ferroelectric phase transition in h-$R$FeO$_3$ film. (b) h-LuFeO$_3$/YSZ and (c) h-YbFeO$_3$/CoFe$_2$O$_4$ interfaces. (d) The thickness-dependent Q. (e) Theoretical phase diagram for paraelectric-ferroelectric phase transition. (f) Thickness-dependent Tc in h-ScFeO$_3$ thin film. (g) The correlation between ferroelectric critical thickness ($\zeta_{FE}$) and effective thickness of interfacial clamping layer ($\zeta_0$). (a) and (e)-(g) are reprinted with permission from ref 53. Copyright 2023, American Physical Society. (b) is reprinted with permission from ref 60. Copyright 2017, American Physical Society. (c) and (d) are reprinted with permission from ref 59. Copyright 2022, American Physical Society.

Therefore, the boundary of phase transition, as shown in Fig. 4(e), is determined by $T_s$ and $\zeta_0$ simultaneously. The phase transition based on in-situ reflection high-energy electron diffraction (RHEED) was observed in h-ScFeO$_3$ films on both STO (111) and Al$_2$O$_3$ (001) substrates, which can be fitted using the above phenomenological model (see Fig. 5(f)). Moreover, this model also indicates that the "practical" critical thickness of h-$R$FeO$_3$ is proportional to the characteristic thickness of interfacial clamping (see Fig. 5(g)). Therefore, to remove the interfacial clamping-induced critical thickness in ultrathin limit of h-$R$FeO$_3$ films, it is required for lattice-matched substrates or freestanding membranes released from substrates.

## 2.4 Strain effects on improper ferroelectricity

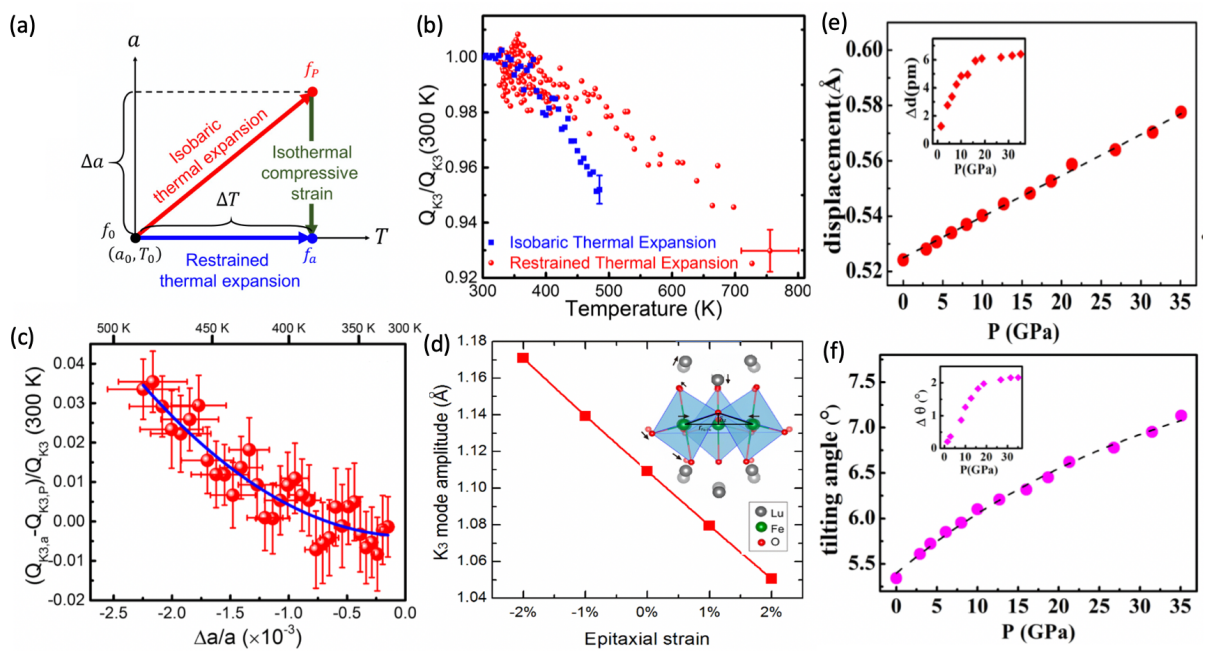

**Fig. 5** (a) Schematic of isobaric and restrained thermal expansion in ($a$, $T$) space. (b) The temperature dependent K$_3$ distortion for isobaric and restrained thermal expansion. (c) The effect of isothermal compressive strain on K$_3$ distortion. (d) The effect of biaxial compressive strain on K$_3$ distortion by calculation. (e) The pressure dependent displacement of rare earth and (f) tilt angle of FeO$_5$. (a) to (d) are reprinted with permission from ref 66. Copyright 2017. American Physical Society. (e) and (f) are reprinted with permission from ref 64. Copyright 2019. American Physical Society.

Strain engineering, as a powerful strategy, could be utilized to stabilize the crystal structure that is unstable in bulk and tune the ordering temperatures and electronic states [61]. Owing to the nature of improper ferroelectrics linked with the K$_3$ distortion, strain engineering serves as a more effective way to modulate improper ferroelectricity in h-$R$FeO$_3$, compared to proper ferroelectrics. Experimentally, the control of strain (gradients) can be realized by varying misfit strain imposed by various substrates [30], chemical doping [31,62], inserting buffer layer [59], changing thicknesses [63], and hydrostatic pressure [64,65].

The misfit strain originates from the lattice mismatch at the interface, which is also susceptible to the temperature due to thermal expansion [66]. For isobaric thermal expansion, both $a$ and $c$ of h-$R$FeO$_3$ increase with temperature. On the other hand, if only the film is heated (substrate remains the same temperature), the in-plan lattice constant ($a$) is locked by the substrate while the out-of-plane lattice constant ($c$) is free to changes, corresponding to restrained thermal expansion. By comparing the isobaric thermal expansion and the restrained thermal expansion, the effect of isothermal strain can be detected, as illustrated in Fig. 5(a). As shown in Fig. 5(b), with the increase of temperature, both the isobaric and restrained thermal expansion could lead to the decrease of $Q$ for K$_3$ distortion, but the change of $Q$ is larger for isobaric thermal expansion. Combining the two types of thermal expansions, we extract the isothermal strain |$\Delta a/a$| and its effect on $Q$, as shown in Fig. 5(c). These results imply that the in-plane compressive strain enhances the K$_3$ distortion. The first-principles calculations corroborate the effect of epitaxial strain, as shown in Fig. 5(d). As illustrated in the inset of Fig. 5(d), under the in-plane compressive strain, the oxygen atom at the center of the trimer, which is shared by three bipyramids, moves up, while the other oxygen atoms in the bases of the three bipyramids all move down, corresponding to enhancement of the K$_3$ distortion. Since the polarization is proportional to $Q^3$ (see eqt. (2)), the in-plane compressive strain is expected to enhance the ferroelectricity polarization.

In addition, the isovalent chemical substitution (replacing cations with the different ionic radius but the same valence) can tune the lattice constant, corresponding to the chemical strain [31,62]. Furthermore, the hydrostatic pressure can modulate strain by reducing both lattice constants $a$ and $c$ and modulate the $c/a$ ratio. Since Fe 3d - O 2p hybridization prefers to be along the $c$ axis, the out-of-plane lattice constant is more stable against pressure. As shown in Fig. 5(e) and (f), based on density function theory (DFT) calculation, as the pressure increases from 0 to 35 Ga, both the displacement in rare earth layer and the tilt angle of FeO$_5$ bipyramids increase, which is verified by the experimental results qualitatively (see insets of Fig.5 (e) and(f)) [64].

## 2.5 Ferroelectric switching dynamics in h-$R$FeO$_3$ films

The ferroelectric switching mechanism in the single crystal of bulk hexagonal manganites is usually described by the Kolmogorov-Avrami-Ishibashi (KAI) model [67], which is strongly correlated with the domain wall motion, wherein the six-fold ferroelectric vortex is pinned at the defect site [68-70]. By contrast, the small grains (~ 10 nm) and the existence of interface in thin-film h-$R$FeO$_3$ favors the switching modes described by the nucleation limited switching (NLS) model [71]. The NSL model treats the polarization switching as inhomogeneous nucleation, as illustrated in Fig. 6(a), with a distribution of nucleation time. The frequency dependent-coercive field can be fitted by both the Ishibashi-Orihara model ($E_C \propto f^\beta$, with $\beta = 0.28$) and the Du-Chen model ($\ln f_0 + \alpha/E_C^2$)[72,73], which are based on the KAI and the NLS models respectively. Moreover, the time-dependent change of polarization can be well fitted by the NLS model (see

Fig. 6 (c) and (d)), indicating that the nucleation process may dominate over the domain wall motion in h-$R$FeO$_3$ films, although the domain wall motion and homogeneous switching cannot be fully excluded [74].

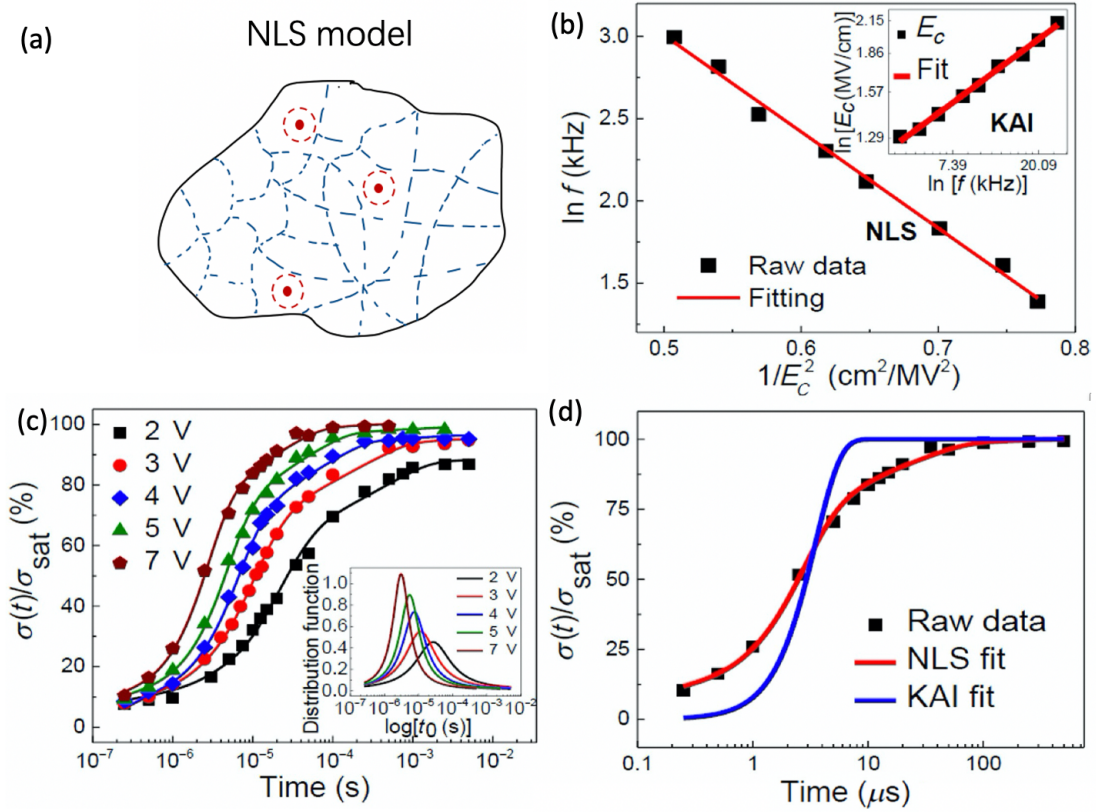

**Fig. 6** (a) Schematic diagram of the NLS model. (b) The frequency-dependent coercive field for h-YbFeO$_3$ film and related theoretical fitting. (c) The time-dependent polarization under different voltage for h-YbFeO$_3$ and (d) related fitting by NLS and KAI model. (b) to (d) are reprinted with permission from ref 59. Copyright 2022. American Physical Society.

## 2.6 Quantification of K$_3$ distortion from STEM image

Since the K$_3$ distortion plays a key role in determining the improper ferroelectricity of h-$R$FeO$_3$, the direct mapping of order parameters $(Q, \phi_Q)$ is essential to reveal the behavior of ferroelectric domains or vortex and examine the proposed model from the Landau theory [51-53]. For h-$R$FeO$_3$, the displacement of rare earth ($u$) can be connected to the K$_3$ distortion as [75,76]:

$$u = u_0 + Q\,'\cos(\phi_Q - \vec{q} \cdot \vec{r}_n) \qquad (7)$$

where the amplitude of rare-earth corrugation ($Q'$) is proportional to $Q$. As shown in Fig. 7(a), viewed along the [100] direction, the atomic position of rare earth ($x_i, y_i$) can be quantified from two-dimensional gaussian fitting of STEM image. Since displacement pattern of rare earth repeats every three atoms for the single domain states, the above equation can be further modified as:

$$u_i = Q'\cos(\phi_Q - \frac{2\pi}{3}\frac{x_i}{a_P}) \tag{8}$$

in which $a_P$ is the in-plane distance between two neighboring $R$, $u_i$ is the relative displacement of the $R$ atom, with respect to average $y_i$ of three $R$ atoms in the unit cell. Based on the least-square fit of the $R$ position using above equation, the order parameter $(Q', \phi_Q)$ can be extracted. Fig. 7(b) shows the schematic patterns of three neighboring $R$ atoms for different ferroelectric domains.

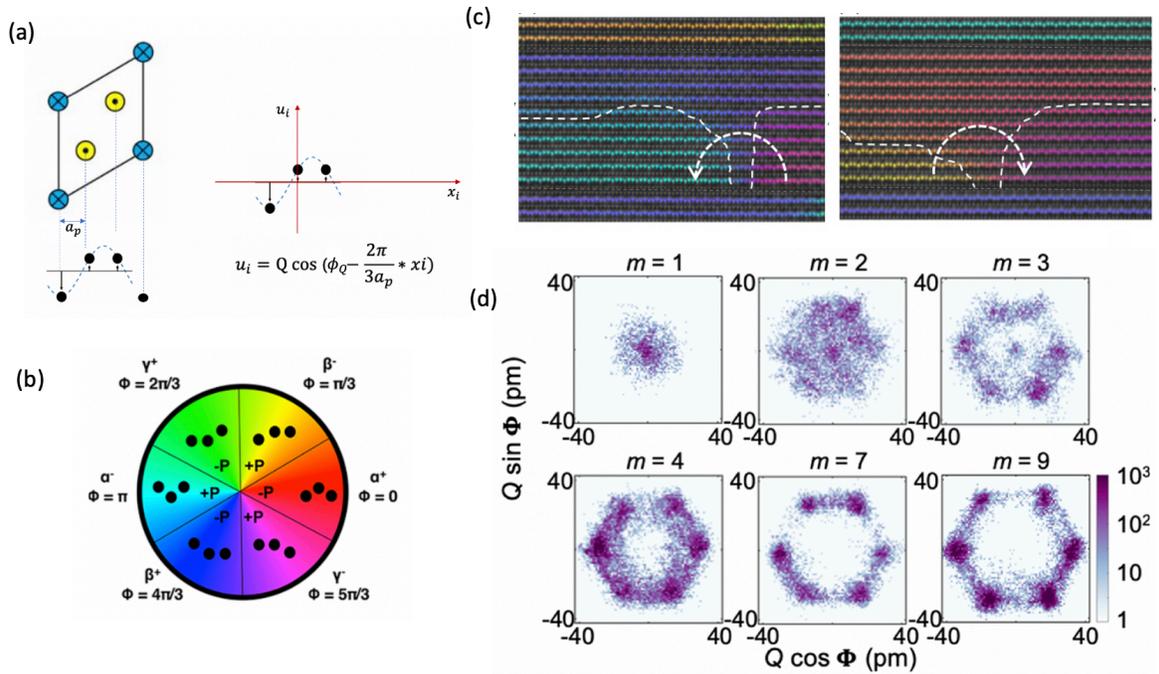

**Fig. 7** (a) Schematics for the quantification of $(Q', \phi_Q)$, based on the positions of rare earth. (b) Patterns of rare earth with different $\phi$. (c) Ferroelectric domain distribution and the overlap of $\phi_Q$ in (h-LuFeO$_3$)$_m$/LuFe$_2$O$_4$ superlattice. (d) $(Q', \phi_Q)$ distributions in (h-LuFeO$_3$)$_m$/LuFe$_2$O$_4$ superlattice with different m. (b) to (d) are reprinted with permission from ref 76. Copyright 2022. American Physical Society.

Moreover, this method can be applied to visualize ferroelectric domain distributions in real space. As shown in Fig.7(c), the half-vortex is pinned at the interface in (h-LuFeO$_3$)$_m$/(LuFe$_2$O$_4$) superlattices [76-78]. By varying *m*, the confinement of electrostatic boundary condition on the topological defects can be quantified from the distribution of the order parameters (see Fig. 7(d)). As the thickness of h-LuFeO$_3$ layer increases, $\phi_Q$ for the majority of domains is $n*\frac{\pi}{3}$ with integer *n*, which is consistent with the energy landscape from the Landau theory. When *m* decreases, an emergent P3c1 symmetry is observed for *m* = 2 before h-LuFeO$_3$ becomes paraelectric phase with *m* = 1 [76,79]. Similar methods have also been applied to track the improper ferroelectricity near phase transition, interface, and defects[80]. These results provide an unprecedented degree of control over ferroelectric domains in hexagonal rare-earth ferrites, pushing the boundary of the fundamental understanding of improper ferroelectrics and the exploration of novel functionalities.

## 3. Magnetism

Beyond the improper ferroelectricity, the structural distortions also play the key role in the spin-lattice coupling by the modulating magnetic free energy [81-83], which provides the microscopic picture to understand the spin orders in h-$R$FeO$_3$, such as Neel temperature ($T_N$), net magnetization of canted antiferromagnetism, and the single-ion magnetic anisotropy.

### 3.1 Spin-lattice coupling in h-$R$FeO$_3$

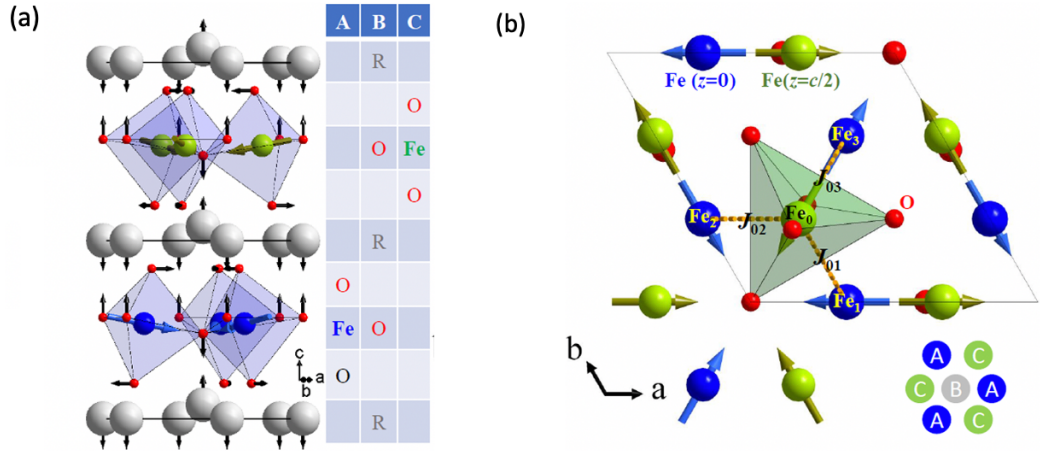

**Fig. 8** (a) Schematic for the canted-antiferromagnetism under the K$_3$ distortion in h-$R$FeO$_3$. (b) The top view for the alignment of Fe spins with A$_2$ magnetic structure and related interlayer exchange interaction. (a) and (b) are reprinted with permission from ref 31. Copyright 2018. American Physical Society.

The Fe spins associated with magnetic order of h-$R$FeO$_3$ are displayed in Fig. 8(a), which are mostly in-plane with a small canting toward the $c$ axis, corresponding to weak ferromagnetism or canted-antiferromagnetism [30-32,84,85]. Within the triangular lattice of FeO$_5$, the in-plane angle between neighboring Fe spins is 120°, corresponding to an antiferromagnetic order, due to a strong exchange interaction within the FeO layer.

The general expression for the effective spin Hamiltonian of a lattice is [33]:

$$H = \sum_{ij} J_{ij} \vec{S}_i \cdot \vec{S}_j + \sum_{ij} \vec{D}_{ij} \cdot (\vec{S}_i \times \vec{S}_j) + \sum_i \vec{S}_i \cdot \overleftrightarrow{\tau}_i \cdot \vec{S}_i \qquad (9)$$

in which $J_{ij}$ is exchange interaction, $\vec{D}_{ij}$ is the DM vector, $\overleftrightarrow{\tau}_i$ is for single-ion anisotropy (SIA) tensor. Therefore, the key for understand spin-lattice coupling is to reveal how the structural distortions influence $J_{ij}$, $\vec{D}_{ij}$, and $\overleftrightarrow{\tau}_i$ in h-$R$FeO$_3$.

The effective inter-layer exchange interaction relies on the K$_3$ lattice distortion. For the inter-layer exchange interaction of a given Fe, there are three nearest Fe spins, as shown in Fig. 8(b). The K$_3$ distortion breaks the three-fold rotational symmetry and makes $J_{03}$ unequal to $J_{02}$ and $J_{01}$, resulting in a non-zero effective interlayer exchange energy [31]:

$$E_{inter} \propto (J_{01} - J_{03})S(S+1). \tag{10}$$

which is expected to determine the three-dimensional ordering and the Neel temperature. Moreover, for FeO$_5$ bipyramids, the electronic structure of Fe is influenced by crystal field of nearby oxygen [86], which is the origin of single-ion anisotropy (SIA).

### 3.2 K$_3$ distortion and Neel temperature ($T_N$) in h-$R$FeO$_3$

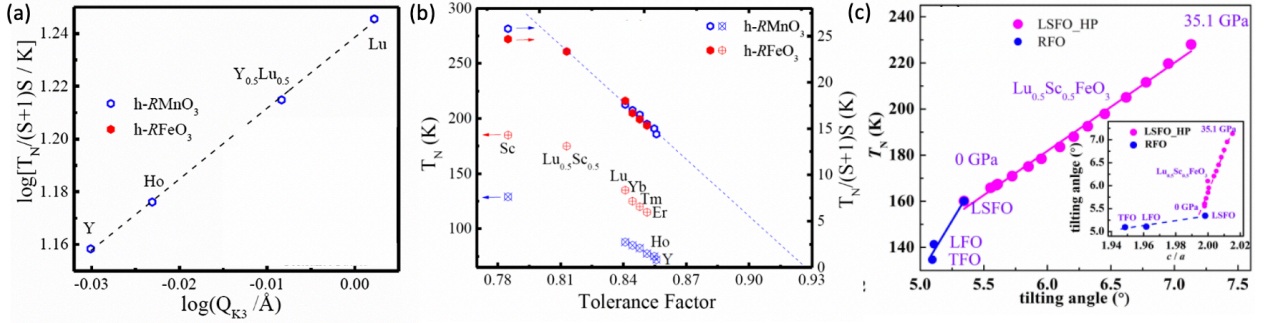

**Fig. 9** (a) log $[T_N/(S+1)S/K]$ as a function of log $(Q)$. (b) The dependence of $T_N$ on tolerance factor. (c) The pressure-dependent tilting angle of FeO$_5$ and $T_N$. (a) and (b) are reprinted with permission from ref 31. Copyright 2018. American Physical Society. (c) is reprinted with permission from ref 64. Copyright 2019. American Physical Society.

Due to the large difference between the magnetic ordering temperature ($T_N$ = 100 - 200 K) and the ferroelectric Curie temperature ($T_C \approx$ 1000 K), $T_N$ is the bottleneck to achieve room-temperature multiferroicity in single-phase h-$R$FeO$_3$. Since strain can modify lattice constant and resultant Fe site positions, the correlation between K$_3$ distortion and $T_N$ through the strain effect has been carefully investigated [30, 31]. The K$_3$ distortion reduces the local symmetry of FeO$_5$ from C$_{3v}$ to C$_s$, which leads to the interlayer exchange interaction $E_{inter}$ in eqt. (10). $T_N$ is expected to be approximately proportional to $E_{inter}$.

Moreover, the relationship between $J_{01} - J_{03}$ and K$_3$ distortion is proposed as [31,33]:

$$J_{01} - J_{03} \propto a_2 Q_{K3}^2 + a_4 Q_{K3}^4 \tag{11}$$

where $a_2$ and $a_4$ are coefficients. As shown in Fig. 9(a), the power law relationship can be identified as:

$$\frac{T_N}{S(S+1)} \propto Q_{K3}^n \qquad (12)$$

with n =2.7 ± 0.05, which is consistent with the Eq. (11). The smaller rare earth atom is expected to reduce the in-plane lattice constant. The tolerance factor $t = (r_R + r_O)/[(r_{TM} + r_O)\sqrt{2}]$ can also be used to gauge the $K_3$ structure distortion in h-$R$FeO$_3$. As shown in Fig. 9(b), by varying the rare earth atom from Er to Sc, the smaller tolerance factor corresponds to stronger $K_3$ distortion Correspondingly, $T_N$ increases from 125 K for h-ErFeO$_3$ to 185 K for h-ScFeO$_3$. A similar trend of $T_N$ also exists in h-RMnO$_3$ [87,88]. On the other hand, the hydrostatic pressure could also modulate $c/a$ ratio and the $K_3$ distortion, the relationship between the $c/a$ ratio and the tilt angle of FeO$_5$ bipyramids is given in the inset of Fig. 9(c). With the increase of hydrostatic pressure, the enhanced $K_3$ distortion leads to the increase of $T_N$ for Lu$_{0.5}$Sc$_{0.5}$FeO$_3$ [64]. Moreover, the electron doping for R site or ion doping in Fe site may also modify the magnetization[89,90].

### 3.3 $K_3$ distortion and weak ferromagnetism in h-$R$FeO$_3$

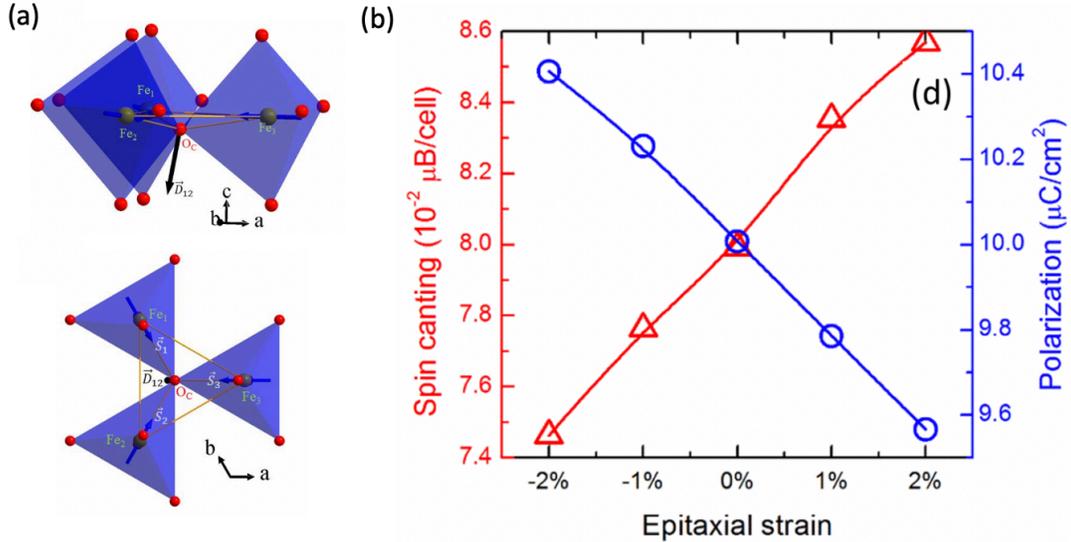

**Fig.10** (a) Schematics of FeO$_5$ trimer from side view and top view. (b) The epitaxial strain-dependent net magnetic moment and polarization in h-$R$FeO$_3$, using DFT. (a) is reprinted with permission from supplementary of ref 31. Copyright 2018. American Physical Society. (b) is reprinted with permission from ref 66. Copyright 2017. American Physical Society.

The net magnetization along the $c$ axis is proportional to the canted angle of Fe spins, which originates from the competition between the intralayer exchange interaction and the DM interaction [91-93]. As shown in Fig. 10(a), the $K_3$ structural distortion causes a displacement of oxygen atom at center ($\delta$) along the $c$ axis. The vector coefficient of DM interaction ($\vec{D}_{12}$) for neighboring Fe$_1$ and Fe$_2$ is perpendicular to the Fe$_1$-O$_c$-Fe$_2$ plane, corresponding to a direction along

$$\left(\frac{a\delta}{\sqrt{3}}, 0, \frac{\sqrt{3}a^2}{18}\right) \quad (13)$$

where $a$ is the in-plane lattice constant. This direction can be represented by a small tilt angle $\phi_{DM}$ away from the $c$ axis with $\tan(\phi_{DM}) = \frac{6\delta}{a}$. The magnitude of the DM interaction follows $D \propto a^2$ approximately because $\delta \ll a$. By minimizing the total energy of exchange interaction and DM interaction, the canting angle of Fe spin follows:

$$\theta_{cant} \approx \frac{D}{\sqrt{3}J}\phi_{DM} \propto \frac{a^2\phi_{DM}}{J} \quad (14)$$

The above model indicates that the in-plane compressive strain cannot decide the net magnetic moment directly. The smaller in-plane lattice constant ($a$) is expected to reduce the DM interaction (because $D \propto a^2$) and enhance the exchange interaction $J$, which decreases $\theta_{cant}$. On the other hand, smaller $a$ can enhance the K$_3$ distortion and the tilt angle $\phi_{DM}$, which increases $\theta_{cant}$. Experimentally, by selecting the rare earth with different atomic radius, it is shown that a reduced $a$ actually leads to a smaller net magnetic moment [31], suggesting that the strain effect on the DM and exchange interactions dominates. Fig. 10(b) demonstrates the relationship between the spin canting and the in-plane epitaxial strain calculated by DFT [66], which is consistent with the experiments qualitatively. Therefore, the unexpected reduction of net magnetic moment under compressive strain in h-$R$FeO$_3$ can be understood based on spin-lattice coupling that has a strong effect on the DM interaction.

### 3.4 Effect of structure distortion of FeO$_5$ on single-ion anisotropy

In h-$R$FeO$_3$, the reversal of weak ferromagnetic moments along the $c$ axis is equivalent to the rotation of the spins by 180° within the basal plane, according to the single-ion magnetic anisotropy. When Fe is displaced with respect to O in FeO$_5$ bipyramids (see Fig. 11(a)), the symmetry of FeO$_5$ bipyramids reduces from D$_{3h}$ to C$_s$, leading to a single-ion magnetic anisotropy via spin-lattice coupling. The one-electron Hamiltonian to describe the single-ion magnetic anisotropy can be written as[45]:

$$H_\alpha = V_{CF}(d) + \xi\vec{L}\cdot\vec{S} - J\vec{S}\cdot\hat{\alpha} \quad (15)$$

where the first term is the energy of Fe electrons in the crystal field, which depends on the perturbation parameter $d$ that is proportional to structural distortion of FeO$_5$ bipyramids ($d \propto -\delta_{Fe}$). The second term is the spin-orbital coupling term and the third term represents molecular field on the Fe spin. The single-ion anisotropy energy can be found from the dependence of total energy on direction $\hat{\alpha}$, with $\hat{\alpha} = \hat{x}, \hat{y}, \hat{z}$, and the anisotropy energy is lower in the $x$ direction ($E_x - E_z < E_y - E_z$) for $\delta_{Fe} > 0$ [45].

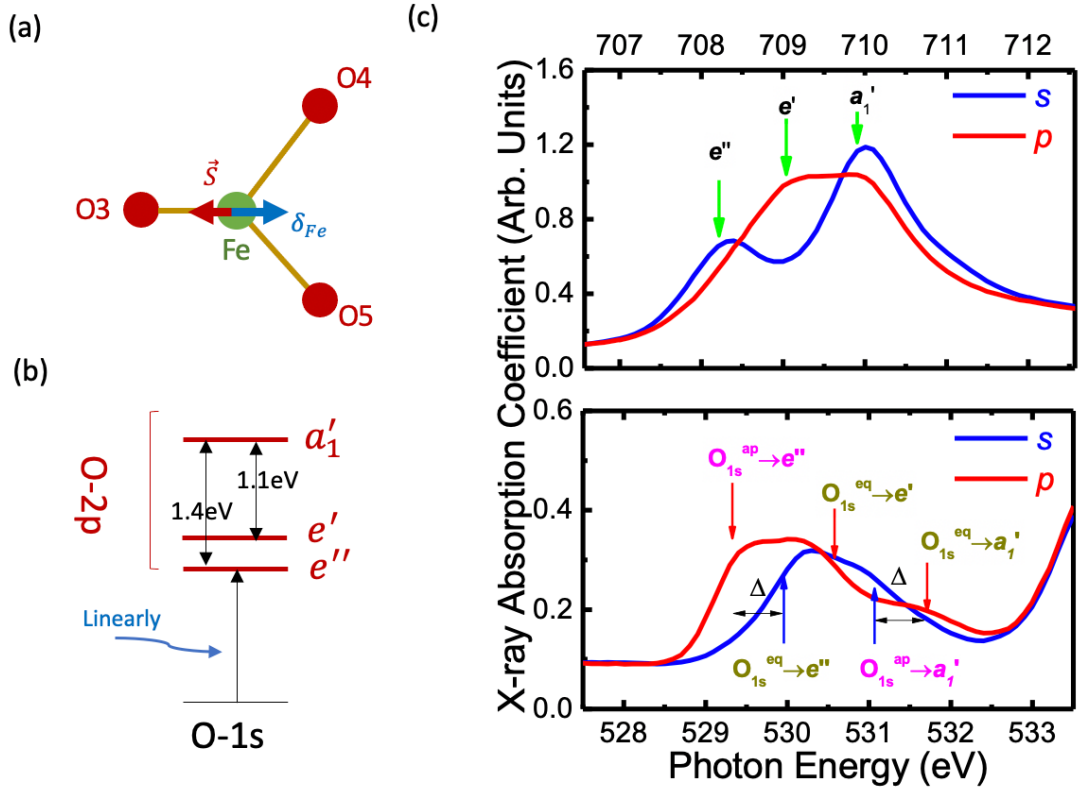

**Fig.11** (a) Schematics of $K_1$ mode distortion within $FeO_5$. (b) Schematic for the XAS of O K edge and (c) related experimental XAS spectra in h-$LuFeO_3$ for Fe $L_3$ edge and O K edge.

For h-$RFeO_3$, one electron picture has been used to interpret the x-ray absorption spectroscopy (XAS) of oxygen-$K$ edge [94,95], since the energy of unoccupied oxygen $p$ states is determined by their hybridization with the Fe 3d states (see Fig. 11(c)). By analyzing the XAS spectra for both Fe $L_3$ and O K edge, the hybridization between Fe 3d and O 2p can be verified. (see Fig. 11(c)). The experimental evidence of hybridization and crystal-field splitting provides critical parameters in Eq. (15) to decide the single-ion anisotropy due to the $K_1$ distortion.

**3.5 The ferrimagnetic order in h-YbFeO$_3$**

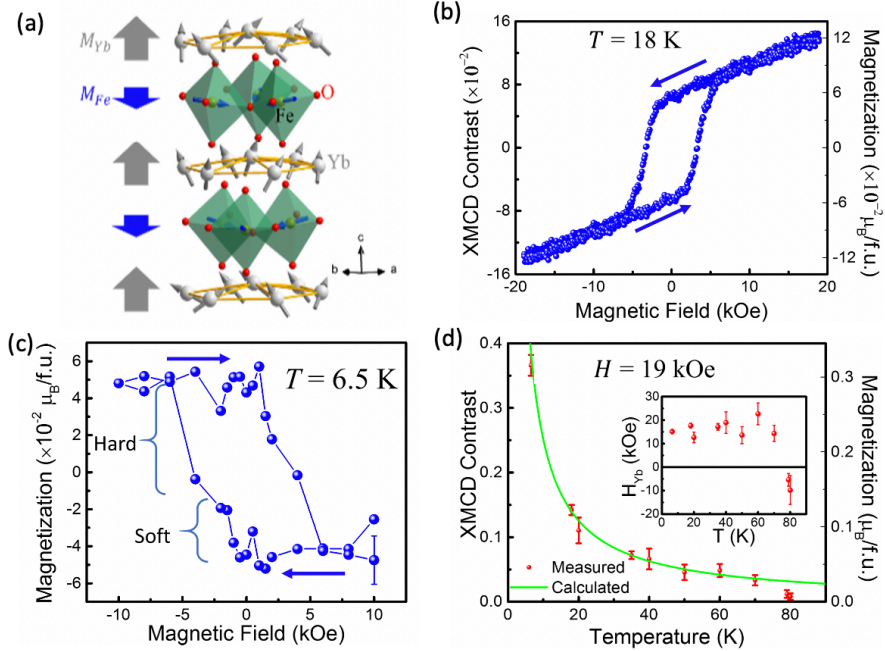

**Fig. 12** (a) Schematics for the spin configuration in h-YbFeO$_3$. (b) The magnetic field-dependent XMCD contrast for Yb M$_5$ edge and related magnetic moment of Yb. (c) Magnetic field dependence of the magnetization of Fe at 6.5 K. (d) The temperature dependence of XMCD contrast for Yb M$_5$ edge. (a) to (d) are reprinted with permission from ref 96. Copyright 2017. American Physical Society.

When $R$ sites are occupied by magnetic ions, such as Yb$^{3+}$, the interplay between the $R$ sublattice and Fe sublattice leads to ferrimagnetic order due to the difference in magnitude of the $R$ and Fe spins [96-98]. A schematic of spin arrangement for Yb and Fe sites are depicted in Fig. 12(a), where the net magnetic moment of the Yb layer is antiparallel with the net magnetic moments of the Fe layer. According to the mean-field theory [99-101], the magnetization of Yb can be approximated as:

$$M_{Yb} = \mu_{Yb}^2 \frac{\Gamma_{YbFe} M_{Fe} + \mu_0 H}{3 k_B T} \quad (16)$$

where $\mu_{Yb}$ is the magnetic moment of Yb, $\Gamma_{YbFe}$ is the molecular field parameters for Yb-Fe interaction, $H$ is the magnetic field, $M_{Fe}$ is the magnetization of the Fe sublattice, $T$ is temperature, $\mu_0$ and $k_B$ are vacuum permeability and Boltzmann constants, respectively. As shown in Fig. 12(b), the x-ray magnetic circular dichroism (XMCD) contrast of Yb M$_5$ edge shows a linear dependence at high field. According to Eq. (16), the slope for the field dependence is $\chi_{Yb} = \frac{\mu_{Yb}^2 \mu_0 H}{3 k_B T}$, from which one can calculate $\mu_{Yb} = 1.6 \pm 0.1\ \mu_B$, which is significantly reduced with respect to the magnetic moment of free Yb (4.5 $\mu_B$). The reduced value of $\mu_{Yb}$ can be attributed to the excitation of the 4$f$ electrons in Yb to low-lying crystal-field states. The hysteresis of Fe magnetization shows a reversed field-dependence (see Fig. 12(c)), indicating that $M_{Fe}$ and $M_{Yb}$ have different sign in the zero field, corresponding to the ferrimagnetic orders and $\Gamma_{YbFe} < 0$. The similar magnetic coercive fields for Fe and Yb also indicate that the exchange field on Yb is generated by the Fe sites.

When temperature is much lower than the magnetic ordering temperature (≈ 120K for h-YbFeO$_3$), $M_{Fe}$ can be treated as constant of the saturation value. Combining the measured $M_{Yb}$ in fixed magnetic field (~ 19 kOe) at different temperature, the exchange field on Yb ($H_{Yb}$) can be calculated using Eq. (16). The result is $H_{Yb} \approx 17$ kOe at low temperature (see inset of Fig. 12 (d)). $H_{Yb}$ changes sign at about 80 K, indicating a possible realignment between the magnetization $M_{Fe}$ and the external field.

### 3.6 Magnetic domains in h-$R$FeO$_3$

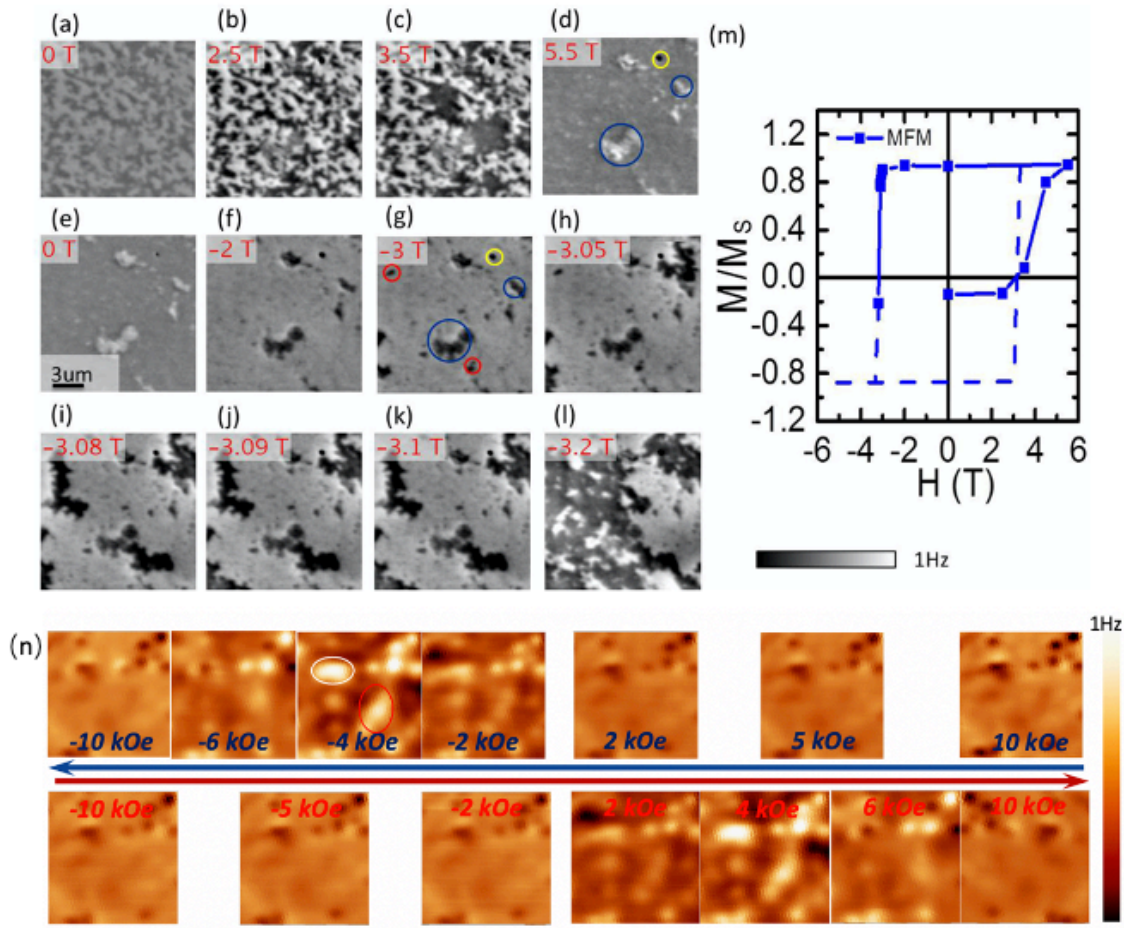

**Fig.13** (a) to (l) MFM images of 200-nm-thick h-LuFeO$_3$/YSZ (111) film measured under different magnetic field after ZFC. (m) The *M-H* curve extracted from the MFM images from (a) to (l). (n) MFM images of h-YbFeO$_3$/YSZ (111) film under different magnetic field, the scan size is 3*3 um. (a) to (m) are reprinted with permission from ref 60. Copyright 2017. American Physical Society. (n) is reprinted with permission from ref 102. Copyright 2023. Springer Nature, under a Creative Commons Attribution 4.0 International License.

Magnetic anisotropy in h-$R$FeO$_3$ is uniaxial because the net magnetic moment comes from the canted antiferromagnetism along the $c$ axis. The small net magnetization results in small magnetostatic energy and large magnetic domains. However, when the $R$ sites are magnetic, like in h-YbFeO$_3$, the larger magnetization and larger magnetostatic energy reduce the domain size. Magnetic domains in h-$R$FeO$_3$ films with and without ferrimagnetic order have been investigated using magnetic force microscopy (MFM) at cryogenic temperature [60,99].

Fig.13 (a-l) show the MFM images at 6 K after zero field cooling (ZFC) for a 200-nm-thick h-LuFeO$_3$/YSZ (111) film [60], in which the magnetization only comes from the Fe spins. The labyrinth-like virgin domain state, with the domain size of ~1.8 um, is identified at zero field (Fig. 13(a)). The domain wall starts moving when the magnetic field reaches about 3.5 T. With increasing the magnetic field, the magnetization saturates at 5 T. The magnitude of magnetization remains constant when the field is decreased from 5 T to 0 T, indicating a perpendicular magnetic anisotropy. As the magnetic field is changed from −3 T to −3.2 T, the up domain is reversed via domain nucleation and domain wall propagation. Fig. 13(m) shows the $M$-$H$ curve deduced from the MFM images, indicating that the h-LuFeO$_3$ is a hard ferromagnet. Moreover, the temperature-dependent MFM images exhibit that the domain contrast becomes weaker with increasing temperature, following a mean-field-like behavior with $T_c \approx 148$K. For h-YbFeO$_3$, a much smaller magnetic domain size on the order of several hundred nanometers was observed [102], due to the stronger magnetostatic contribution from Yb, as shown in Fig. 13(n). The enhanced domain contrast can be attributed to nucleation of magnetic domains when the magnetic field is close to the magnetic coercive filed (approximately ±4000 Oe).

## 4. Magnetoelectric couplings

In single-phase multiferroic h-$R$FeO$_3$, the improper nature of ferroelectricity as well as the spin-lattice coupling imply that the structural distortions may serve as the bridge for the interplay between spin order and electrical polar order, leading to a magnetoelectric (ME) coupling [30-33,103-105]. Due to the existence of domains as well as domain walls for both ferroelectricity and magnetism, the ME coupling in h-$R$FeO$_3$ can be divided into the bulk-state ME coupling as well as domain-wall ME coupling.

### 4.1 Magnetoelectric coupling in bulk-state

For bulk or single domain state h-$R$FeO$_3$, we describe the spin direction of Fe and the apical displacement of FeO$_5$ (from the K$_3$ distortion) using $\phi_L$ and $\phi_Q$ respectively. The net magnetization (M) can be described as [51]:

$$M = -M_S \cos(\phi_Q - \phi_L) \qquad (17)$$

Fig. 14(a) gives the correlation of polarization ($P$) and $M$ in triangle-bipyramids of h-RFeO$_3$. As shown in Fig. 14(b), under the external electric field, the switching of polarization direction is accompanied by a rotation of $\phi_Q$ by 60°. If $\phi_L$ does not follow the change of $\phi_Q$, the magnetization direction would reverse, corresponding to the ME coupling for the bulk-state. Otherwise, if $\phi_L$ follows $\phi_Q$, the direction of magnetization will not change, corresponding to the ME decoupling. The energy landscape in ($\phi_Q, \phi_L$) space is given in Fig. 14(c), where the energy barrier is higher on the path of the ME coupling. Therefore, when the polarization is switched by the electric field, the net magnetization is expected to be decoupled from it.

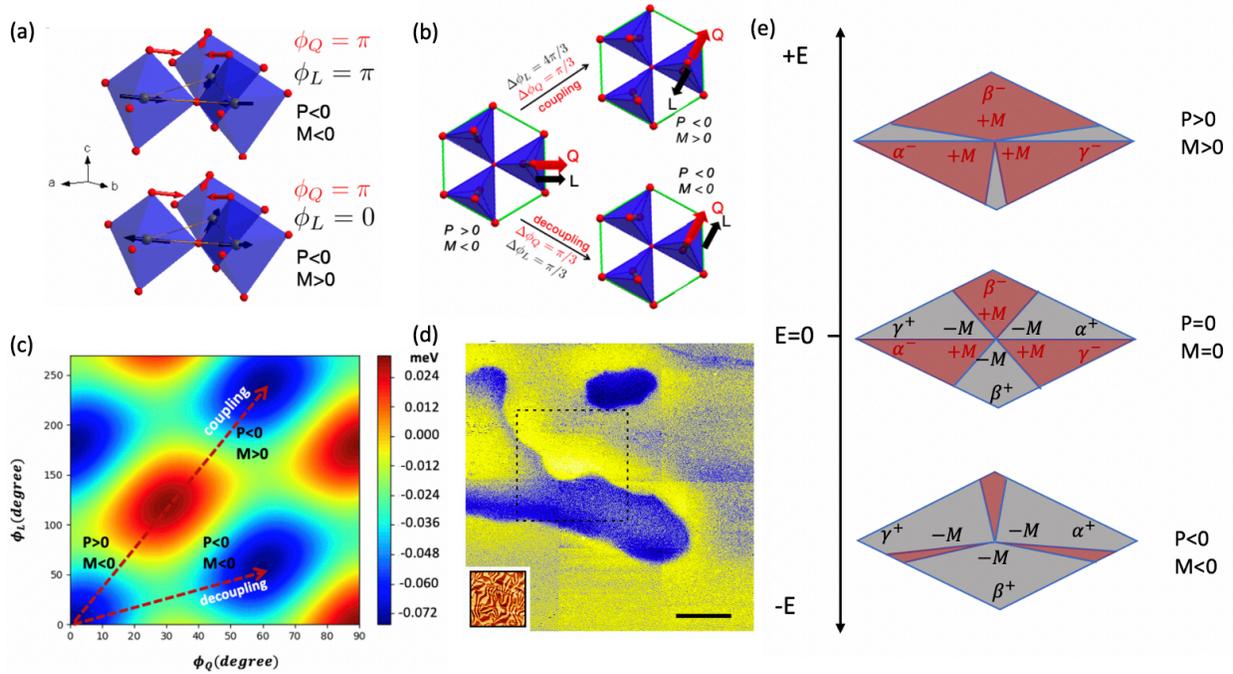

**Fig.14** (a) Schematic for the spin direction ($\phi_L$) and structural distortion ($\phi_Q$) within the triangle lattice of FeO$_5$ bipyramids. (b) Schematic for the ME coupling and decoupling when the polarization is switched from $P > 0$ to $P < 0$. (c) The energy landscape in ($\phi_Q, \phi_L$) space. (d) MFM image of the h-Lu$_{0.5}$Sc$_{0.5}$FeO$_3$, the inset is the PFM phase image. (e) Proposed ME coupling at ferroelectric vortex. (a) to (c) are reprinted with permission from ref 102. Copyright 2023. Springer Nature, under a Creative Commons Attribution 4.0 International License. (d) is reprinted with permission from ref.106. Copyright 2018, Springer Nature, under a Creative Commons Attribution 4.0 International License.

Experimentally, for h-Lu$_{0.5}$S$_{0.5}$cFeO$_3$ single crystals [106], the size of the magnetic domains is on the order of 100 um, while the size of the ferroelectric domain is on the order of 1 um. No identifiable correlations between the magnetic and ferroelectric domains can be observed experimentally. The large domain-size mismatch also suggests the ME decoupling, as shown in Fig. 14(d). Therefore, the bulk-state ME coupling previously predicted [33], in which the magnetization switching follows with the polarization switching at the ferroelectric vortex under external electric filed, as illustrated in Fig. 14(e), may not be the most possible scenario in h-*R*FeO$_3$.

Here, we briefly compare h-RFeO$_3$ with the other widely studied multiferroics, BiFeO$_3$. The bulk-state magnetoelectric coupling in single-phase BiFeO$_3$ hinges on the rotation of magnetic easy plane associating with the 71° or 109° rotation of polarization, which is along the [111] direction in pseudo-cubic structure[14, 107]. Due to the polarization of h-RFeO$_3$ can only be switched along c-axis, as well as the high-energy barrier on switching path ME coupling, the forbidden electric field-driven magnetization switching in h-RFeO$_3$ resembles the similar forbidden behavior in BiFeO$_3$ during 180° polarization switching.

On the other hand, the heterostructures of ferromagnet on BiFeO$_3$ indicate the change of antiferromagnetic easy plane during polarization switching can also couple with neighboring ferromagnetic layers, achieving the strong ME coupling in composite form[108,109]. For h-RFeO$_3$, the known explored magnetic layer is LuFe$_2$O$_4$ and inverse-spinel CoFe$_2$O$_4$. The magnetic Curie temperature is enhanced for the former[77]. The magnetic properties CoFe$_2$O$_4$ layer is still less explored, the potential electric-field driven change of exchange bias may rely on the interfacial monolayer reconstruction.[110]

**4.2 Magnetoelectric coupling in clamped antiferromagnetic domain wall**

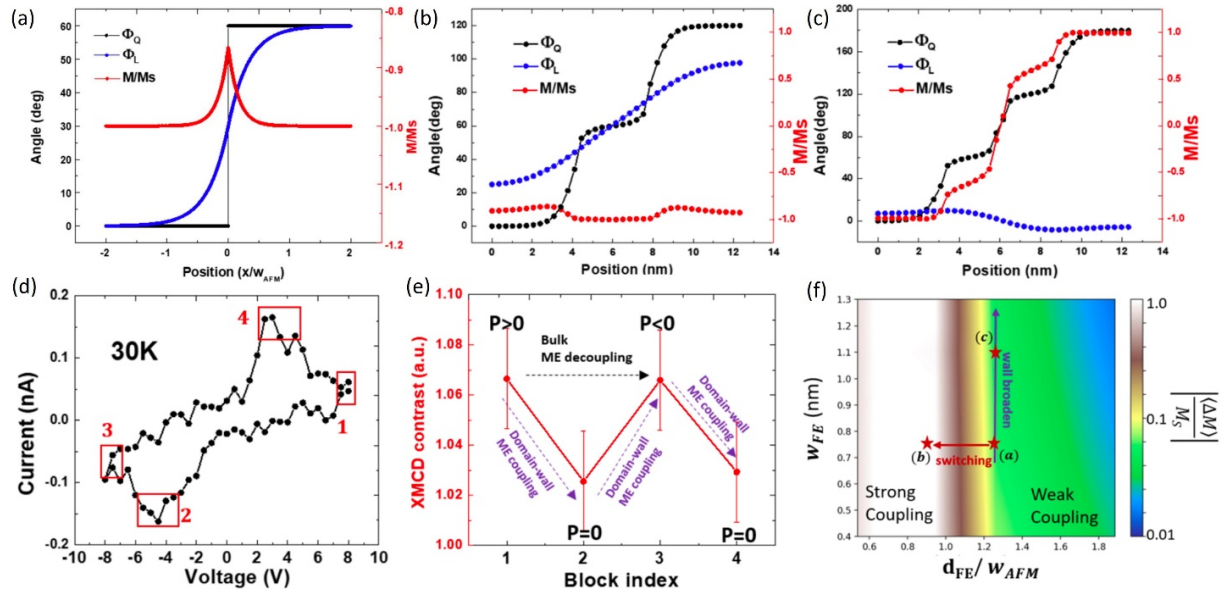

**Fig. 15** (a) Schematic for the single clamped antiferromagnetic (AFM) domain wall in h-YbFeO$_3$. Clamped AFM wall for 12-nm-diameter grains with (b) two and (c) three uniformly distributed ferroelectric domain walls. (d) The voltage-dependent switching current for h-YbFeO$_3$ thin films at 30K. (e) The XMCD contrast of Yb M$_5$ edge for different blocks. (f) Phase diagram of average magnetization reduction due to domain-wall ME coupling, with respect to w$_{FE}$ and d$_{FE}$/w$_{AFM}$. (a) to (f) are reprinted with permission from ref 102. Copyright 2023, Springer Nature, under a Creative Commons Attribution 4.0 International License.

Despite the bulk-state ME decoupling, the ferroelectric domain walls of h-$R$FeO$_3$ could support the ME coupling [111, 112]. As shown in Fig. 15(a), the FE domain wall of h-$R$FeO$_3$ is defined by a sharp change of $\phi_Q$ [27,112], which induces the change of $\phi_L$ due to the single-ion magnetic anisotropy, corresponding to the so-called clamped antiferromagnetic (AFM) domain wall [50]. The magnitude of the magnetization decreases within the clamped AFM domain wall. Moreover, the relatively smaller domain-wall energy in h-$R$FeO$_3$ suggests the possibility of large population of ferroelectric domain walls. By changing electrostatic conditions or grain size in h-$R$FeO$_3$ thin films, the population of FE domain walls can be tuned [113,114] and then influences the distribution of clamped AFM walls (see Fig. 15(b) and (c)). Due to the appearance of FE domain walls, $\phi_L$ cannot follow the change of $\phi_Q$, and thus the net magnetization will be suppressed as the FE domain-wall density increases, as shown in Fig. 15(c). In other words, by controlling the density of FE walls during the ferroelectric switching, the magnitude of the magnetization can be changed, which provides a new ME coupling mechanism linked with ferroelectric and magnetic domain walls.

Experimentally, the polarization switching has been achieved in h-YbFeO$_3$/CFO/LSMO thin films [59], as shown in Fig. 15(d). Blocks 1 and 3 correspond to up and down polarization, while blocks 2 and 4 correspond to zero polarization when the density of FE domain walls is expected to be maximum. Based on the in-situ measurement of XMCD contrast of Yb M5 edge during ferroelectric switching (see Fig. 15(e)), the lower XMCD contrasts in blocks 2 and 4, compared with that of the blocks 1 and 3, substantiate the domain-wall ME coupling discussed above. The critical condition for achieving the domain-wall ME coupling is that the FE domain size ($d_{FE}$) is smaller than the width of the AFM domain wall ($w_{AFM}$), as shown in Fig. 15(f).

## 5. Outlook and summary

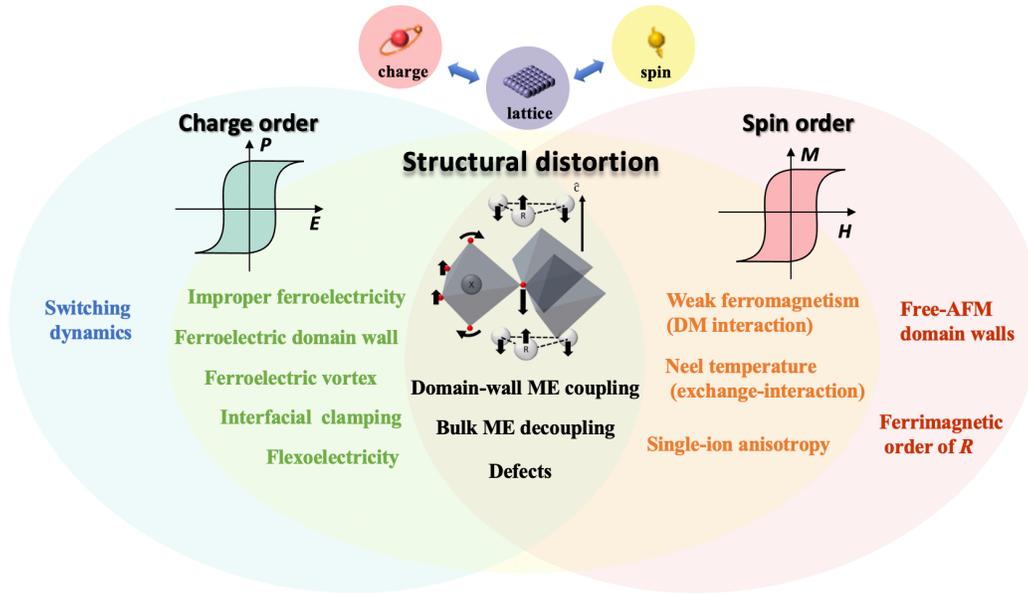

**Fig. 16** The global view of the relation between spin, charge, and lattice of h-RFeO$_3$, representing current understandings.

During the past decades, substantial progresses have been made on atomic-scale synthesis control, high-precision characterization, and multiscale material modeling on hexagonal rare-earth ferrites. This boosts the physic understanding of their fundamental multiferroic properties, particularly the role of non-polar structure distortion on ferroelectricity, magnetism, and ME couplings. The improper ferroelectricity and spin-lattice coupling have been revealed as the origin of the abundant multiferroic properties in h-$R$FeO$_3$, as shown in Fig.16, which is expected to stimulate the fresh ideas for the non-perovskite-structure multiferroics, including but not limiting to:

1. The improper ferroelectric h-$R$FeO$_3$, comparing with their proper perovskite counterparts [115,116], exhibits exotic functionalities, including the dominating role of non-polar structural K$_3$ distortion on ferroic orders, smaller domain wall energy, and the absence of critical thickness. The removal of interfacial clamping (see Section 2.3) would make h-RFeO$_3$ ferroelectric with 1 nanometer thickness, comparable to two-dimensional ferroelectrics.

2. For magnetism, $T_N$ and weak ferromagnetism are determined by the K$_3$ distortion, while the single-ion anisotropy hinges on the K$_1$ distortion. So, methods that can modulate atomic-scale structural distortion, such as chemical doping and hydrostatic pressure, can be effective in further tuning the magnetic properties.

3. Although the bulk-state ME coupling is not favored, the reduced symmetry at the ferroelectric domain walls enables the macroscopic identification for domain-wall ME coupling at nanoscale, which was observed during ferroelectric switching. Hence, ME coupling could be explored in local, low symmetry configurations.

4. Hexagonal rare-earth ferrites h-$R$FeO$_3$ can serve as robust ultrathin ferroelectric template to induce potential proximity-like effect in neighboring materials, such as enhancement of $T_c$ in LuFe$_2$O$_4$[77] and enhancement of spin-Hall angle in heavy metal by the Rashba-Edelstein effect (REE)[117], or control of magnetic skyrmions[89].

Some challenges about multiferroic h-$R$FeO$_3$, however, remain to be addressed in the future, calling for concerted efforts and advanced capabilities in thin-film synthesis, characterization, and modeling. We summarize several challenges and perspectives as follows:

From the perspective of improper ferroelectricity, the demonstration of absence of critical thickness in h-$R$FeO$_3$ has been hindered by the "practical" critical thickness, i.e., the suppression of the K$_3$ distortion at the film/substrate interface due to the structural clamping effect. Other factors, such as oxygen-off stoichiometry [118], interfacial reconstruction [119], should be further clarified, in order to make h-$R$FeO$_3$ scale-free ferroelectric materials experimentally. Moreover, novel electric properties of h-$R$FeO$_3$ are much less explored. For example, the multi-parameter energy diagram and the resulting complex polarization switching path of h-$R$FeO$_3$ may lead to non-traditional polarization-voltage relation under different restrictions [74]. In particular, due to the nature of improper ferroelectricity, the charged domain walls are prevalent[76,112], which may exhibit distinguished electrical performance, in concert with hexagonal structure. The unique response of K$_3$ structural distortion to strain may offer novel mechanism of coupling strain gradient and polarization, i.e., flexoelectricity [120-122]. To address these issues, the synthesis of high-quality hexagonal rare-earth ferrites films is of particular importance. Nevertheless, this is still restricted by the availability of substrates and appropriate electrodes. Expanding the range of available conducting substrates with matching structures is a key step. In turn, for magnetism, questions remain on how to increase $T_N$ of single-phase h-$R$FeO$_3$ film above room temperature, by e.g., chemical doping, epitaxial strain.

Besides the "bulk" state properties, local topological phenomena in h-$R$FeO$_3$ are also important in future research directions. A good example is the famous six-fold ferroelectric vortex which has been demonstrated in single crystal samples as a topological defect with a micrometer scale, accompanied by collective scaling behavior [123-127]. On a smaller scale, screw dislocations, which are often observed in hexagonal materials [128-130], can introduce stronger modulation to the multiferroic properties, and generate topological phenomenon like polar and magnetic skyrmions in h-$R$FeO$_3$ films [131-133].

Regarding substrate clamping effects and limited flexibility, a promising mean to expand strain control is to release freestanding membranes from the substrates and transfer them to various substrates. For example, the compatibility of h-$R$FeO$_3$ on classic perovskite-structure bottom electrode layer, such as h-ScFeO$_3$ on La$_{0.67}$Sr$_{0.33}$MnO$_3$ (LSMO) /SrTiO$_3$(111), as well as atomic-scale synthesis control through *in-situ* monitoring of RHEED, provide an opportunity to achieve freestanding h-$R$FeO$_3$ by dissolving LSMO layer or inserting solvable buffer layer [134,135]. This holds great promise for a larger degree of control over ferroic orders and thus realizing profuse novel phenomena. In addition, the removal of interfacial clamping in h-$R$FeO$_3$ may enhance the polarization state with the thickness as low as monolayer [136-138], due to the improper nature of ferroelectricity.

On application level, the improper ferroelectricity of h-$R$FeO$_3$ shows the great advantages in non-volatile ferroelectric memory devices, such as ferroelectric capacitor, FeRAM, FeFET, which all hinges on the maintenance of remnant polarization after removing electric field[139-141]. Moreover, the thickness-dependent coercivity basically follows the JKD-scaling with Ec ~ t$^{-2/3}$ down to ~ 5nm or even thinner[59], indicating the continuous decrease of coercive volage with shrinking the thickness, which is beneficial for the energy efficiency. Meanwhile, the polarization could maintain above 2 uC/cm$^2$ in ultrathin film. On the other hand, the improper nature of h-$R$FeO$_3$ could also help verify the generality of proposed mechanism and application, based on polarization switching in proper ferroelectrics, such as negative capacitance[74]. However, for the real application of h-$R$FeO$_3$ as ferroelectrics, it still needs to overcome compatibility problem with known semiconductor materials, such as Si, which is less explored to our best knowledge. The domain wall of h-$R$FeO$_3$ is similar to the isomorphic h-$R$MnO$_3$, related anisotropic conductance makes domain walls themself as the electronic components.[142] Recently, the enhanced photovoltaic efficiency in h-$R$FeO$_3$ films via strain engineering also indicates the application potential for photovoltaics devices[143].

Overall, the abundance of emerging phenomena in h-$R$FeO$_3$ have been connected via the complex crystal structure and the spin-lattice coupling one after another. At this point, hexagonal rare-earth ferrites are poised to be exploited for applications in terms of their advantageous electric properties associated with the improper ferroelectricity, particularly in the ultrathin films. In addition, exploring their local topological polar and magnetic properties will be intriguing to understand quantum phenomena in complex materials. We hope that this progress review article may inspire further investigations into hexagonal rare-earth ferrites and their potential applications in future electronics.

**Acknowledgment**

The authors acknowledge the primary support from the National Science Foundation (NSF) through EPSCoR RII Track-1: Emergent Quantum Materials and Technologies (EQUATE), Award No. OIA-2044049. The work was also supported in part by the Nebraska Center for Energy Sciences Research (NCESR).

# Reference:

**1.Introduction:**

1. Fiebig, M., Lottermoser, T., Meier, D. et al. The evolution of multiferroics. Nat Rev Mater1, 16046 (2016).

2. Lu, C., Wu, M. Lin, L., Liu, J-M. Single-phase multiferroics: new materials, phenomena, and physics. National Science Review 6: 653–668 (2019).

3. Ekhard K.H. Salje. Ferroelastic Materials. Annu. Rev. Mater. Res. 42,1.1–1.19(2012).

4. M. Dawber, K. M. Rabe, and J. F. Scott. Physics of thin-film ferroelectric oxides. Rev. Mod. Phys. 77, 1083 (2005).

5. Ramesh, R., Spaldin, N. Multiferroics: progress and prospects in thin films. *Nature Mater* **6**, 21–29 (2007).

6. Spaldin, N.A.,. Why Are There so Few Magnetic Ferroelectrics ? J. Phys. Chem. B 104, 29, 6694–6709 (2000).

7. Spaldin, N.A., Ramesh, R. Advances in magnetoelectric multiferroics. Nature Mater 18, 203–212 (2019).

8. Manfred Fiebig. Revival of the magnetoelectric effect. J. Phys. D: Appl. Phys. 38 R123 (2005).

9. Scott, J. Room-temperature multiferroic magnetoelectrics. NPG Asia Mater 5, e72 (2013).

10. Eerenstein, W., Mathur, N. & Scott, J. Multiferroic and magnetoelectric materials. Nature 442, 759–765 (2006).

11. J, Wang, J. B. Neaton, Zheng, H. et al. Epitaxial $BiFeO_3$ Multiferroic Thin Film Heterostructures. Science. 299, 5613, 1719-1722 (2003).

12. Fiebig, M., Lottermoser, T., Fröhlich, D. et al. Observation of coupled magnetic and electric domains. Nature 419, 818–820 (2002).

13. Valencia, S., Crassous, A., Bocher, L. et al. Interface-induced room-temperature multiferroicity in $BaTiO_3$. Nature Mater 10, 753–758 (2011).

14. J. Wang, JB Neaton, H Zheng, V Nagarajan, SB Ogale, B Liu, D Viehland, V Vaithyanathan, DG Schlom, UV Waghmare, NA Spaldin, KM Rabe, M Wuttig, R Ramesh. Epitaxial $BiFeO_3$ multiferroic thin film heterostructures. Science 299, 5613,1719-172214 (2003).

15. A. Gruverman, D. Wu, H-J Fan, I. Vrejoiu, M. Alexe, R. J. Harrison and J F Scott. Vortex ferroelectric domains. J. Phys.: Condens. Matter 20 342201(2008).

16. Balke, N., Winchester, B., Ren, W. et al. Enhanced electric conductivity at ferroelectric vortex cores in $BiFeO_3$. Nature Phys 8, 81–88 (2012).

17. Das, S., Tang, Y.L., Hong, Z. et al. Observation of room-temperature polar skyrmions. Nature 568, 368–372 (2019).

18. Meier, D., Seidel, J., Cano, A. et al. Anisotropic conductance at improper ferroelectric domain walls. Nature Mater 11, 284–288 (2012).

19. Mundy, J., Schaab, J., Kumagai, Y. et al. Functional electronic inversion layers at ferroelectric domain walls. Nature Mater 16, 622–627 (2017).

20. Meier, D. Functional domain walls in multiferroics. J. Phys.: Condens. Matter 27 463003(2015)

21. J. Seidel, P. Maksymovych, Y. Batra, A. Katan, S.-Y. Yang, Q. He, A. P. Baddorf, S. V. Kalinin, C.-H. Yang, J.-C. Yang, Y.-H. Chu, E. K. H. Salje, H. Wormeester, M. Salmeron, and R. Ramesh. Domain Wall Conductivity in La-Doped $BiFeO_3$. Phys. Rev. Lett. 105, 197603 (2010)

22. G Lawes and G Srinivasan. Introduction to magnetoelectric coupling and multiferroic films. J. Phys. D: Appl. Phys. 44 243001(2011)

23. Martin, L., Rappe, A. Thin-film ferroelectric materials and their applications. Nat Rev Mater 2, 16087 (2017).

24. LW Martin, SP Crane et al. Multiferroics and magnetoelectrics: thin films and nanostructures. J. Phys.: Condens. Matter 20 434220 (2008).

25. Morgan Trassin. Low energy consumption spintronics using multiferroic heterostructures. J. Phys.: Condens. Matter 28 033001(2016).

26. D Khomskii. Classifying multiferroics: Mechanisms and effects. Physics 2: 20 (2009).

27. Choi, T., Horibe, Y., Yi, H. et al. Insulating interlocked ferroelectric and structural antiphase domain walls in multiferroic $YMnO_3$. Nature Mater 9, 253–258 (2010).

28. A. V. Goltsev, R. V. Pisarev, Th. Lottermoser, and M. Fiebig. Structure and Interaction of Antiferromagnetic Domain Walls in Hexagonal $YMnO_3$. Phys. Rev. Lett. 90, 177204 (2003).

29. Van Aken, B., Palstra, T., Filippetti, A. et al. The origin of ferroelectricity in magnetoelectric $YMnO_3$. Nature Mater 3, 164–170 (2004).

30. Wang, W., Zhao, Jun., Wang, W., et al. Room-Temperature Multiferroic Hexagonal $LuFeO_3$ Films. Phys. Rev. Lett. 110, 237601 (2013).

31. Sinha, Kishan., Wang, H., Wang, X., Tuning the Néel Temperature of Hexagonal Ferrites by Structural Distortion. Phys. Rev. Lett. 121, 237203 (2018).

32. Steven M. Disseler, Julie A. Borchers, Charles M. Brooks, et al. Magnetic Structure and Ordering of Multiferroic Hexagonal $LuFeO_3$. Phys. Rev. Lett. 114, 217602 (2015).

33. Das, H., Wysocki, A., Geng, Y. et al. Bulk magnetoelectricity in the hexagonal manganites and ferrites. Nat Commun 5, 2998 (2014).

34. Yang Zhang, Wenlong Si, Yanli Jia, Pu Yu, Rong Yu, Jing Zhu. Controlling Strain Relaxation by Interface Design in Highly Lattice-Mismatched Heterostructure. Nano Lett. 21, 16, 6867–6874 (2021).

35. Johanna Nordlander, Marta D. Rossell, Marco Campanini, Manfred Fiebig, and Morgan Trassin. Inversion-Symmetry Engineering in Layered Oxide Thin Films. Nano Lett.21, 2780−2785 (2021)

36. J. Nordlander, M. D. Rossell, M. Campanini, M. Fiebig, and M. Trassin. Epitaxial integration of improper ferroelectric hexagonal YMnO3 thin films in heterostructures. Phys. Rev. Materials 4, 124403 (2020).

37. Xiaozhe Zhang, Yuewei Yin, Sen Yang, Zhimao Yang, Xiaoshan Xu. Effect of interface on epitaxy and magnetism in h-RFeO3/Fe3O4/Al2O3 films (R= Lu, Yb). J. Phys.: Condens. Matter 29 164001(2017)

38. Shi Cao et al. The stability and surface termination of hexagonal LuFeO3. J. Phys.: Condens. Matter 27 175004 (2015).

39. Jarrett A Moyer, Rajiv Misra, Julia A Mundy, Charles M Brooks, John T Heron, David A Muller, Darrell G Schlom, Peter Schiffer. Intrinsic magnetic properties of hexagonal LuFeO3 and the effects of nonstoichiometry. APL Mater. 2, 012106 (2014)

40. Th. Lonkai, D. G. Tomuta, U. Amann, J. Ihringer, R. W. A. Hendrikx, D. M. Többens, and J. A. Mydosh. Development of the high-temperature phase of hexagonal manganites. Phys. Rev. B 69, 134108 (2004)

41. Hongwei Wang, Igor V. Solovyev, Wenbin Wang, Xiao Wang, Philip J. Ryan, David J. Keavney, Jong-Woo Kim, Thomas Z. Ward, Leyi Zhu, Jian Shen, X. M. Cheng, Lixin He, Xiaoshan Xu, and Xifan Wu. Structural and electronic origin of the magnetic structures in hexagonal LuFeO3. Phys. Rev. B 90, 014436 (2014)

42. Lilienblum, M., Lottermoser, T., Manz, S. et al. Ferroelectricity in the multiferroic hexagonal manganites. Nature Phys 11, 1070–1073 (2015).

43. Th. Lonkai, D. G. Tomuta, U. Amann, J. Ihringer, R. W. A. Hendrikx, D. M. Többens, and J. A. Mydosh. Development of the high-temperature phase of hexagonal manganites. Phys. Rev. B 69, 134108 (2004).

44. A. P Levanyuk, D. G Sannikov. Improper ferroelectrics. Sov. Phys. Usp. 17 199 (1974).

45.Shi Cao et al. On the structural origin of the single-ion magnetic anisotropy in LuFeO3. J. Phys.: Condens. Matter 28 156001 (2016).

46. Xu, X., Wang., W. Multiferroic hexagonal ferrites (h-RFeO3, R=Y, Dy-Lu): a brief experimental review. Modern Physics Letters 28, 21, 1430008 (2014).

47. R. C. Rai, C. Horvatits; D. Mckenna, J. Du Hart. Structural studies and physical properties of hexagonal-YbFeO3 thin films. AIP Advances 9, 015019 (2019).

48. J. Nordlander, M.A .Anderson, C.M. Brooks, M.E. Holtz, J.A. Mundy. Epitaxy of hexagonal ABO3 quantum materials. Appl. Phys. Rev. 9, 031309 (2022).

49. Menglei Li, Hengxin Tan and Wenhui Duan. Hexagonal rare-earth manganites and ferrites: a review of improper ferroelectricity, magnetoelectric coupling, and unusual domain walls. Phys. Chem. Chem. Phys. 22, 14415-14432 (2020).

## 2. Improper ferroelectricity

50. Craig J. Fennie and Karin M. Rabe. Ferroelectric transition in YMnO3 from first principles. Phys. Rev. B 72, 100103(R) (2005).

**2.1 Landau theory for improper ferroelectricity in h-RFeO3**

51. Artyukhin, S., Delaney, K., Spaldin, N. et al. Landau theory of topological defects in multiferroic hexagonal manganites. Nature Mater 13, 42–49 (2014).

52. C. X. Zhang, K. L. Yang, P. Jia, et al. Effects of temperature and electric field on order parameters in ferroelectric hexagonal manganites. J. Appl. Phys. 123, 094102 (2018).

53. Li, X., Yun, Y., Xu, X. S. Improper ferroelectricity in ultrathin hexagonal ferrites films. Appl. Phys. Lett. 122, 182901 (2023)

**2.2 Effect of depolarization field on improper ferroelectricity in h-RFeO3 films**

54. Na Sai, Craig J. Fennie, and Alexander A. Demkov. Absence of Critical Thickness in an Ultrathin Improper Ferroelectric Film. Phys. Rev. Lett. 102, 107601 (2009).

55. Nordlander, J., Campanini, M., Rossell, M.D. et al. The ultrathin limit of improper ferroelectricity. Nat Commun 10, 5591 (2019).

56. D. D. Fong, G. B. Stephenson, S. K. Streiffer, J. A. Eastman, O. Auciello, P. H. Fuoss, and C. Thompson, Ferroelectricity in Ultrathin Perovskite Films., Science 304, 1650 (2004).

57. Javier Junquera and P. Ghosez, Critical Thickness for Ferroelectricity in Perovskite Ultrathin Film, Nature 422, 506 (2003).

58. Y. S. Kim et al., Critical Thickness of Ultrathin Ferroelectric BaTiO3 Films, Appl Phys Lett 86, 102907 (2005).

59. Yu Yun et al. Spontaneous Polarization in an Ultrathin Improper-Ferroelectric/Dielectric Bilayer in a Capacitor Structure at Cryogenic Temperatures. Phys. Rev. Applied 18, 034071 (2022).

**2.3 Interfacial clamping on critical thickness in ultrathin h-RFeO3 film**

60. Wang. W. et al. Visualizing weak ferromagnetic domains in multiferroic hexagonal ferrite thin film. Phys. Rev. B 95, 134443 (2017).

**2.4 Effect of strain on improper ferroelectricity**

61. R. J Zeches, M. D Rossell, J. X . Zhang et al. A Strain-Driven Morphotropic Phase Boundary in BiFeO3. Science 326, 977-980 (2009)

62. Shiqing Deng, Jun Li, Didrik R. Småbråten, Shoudong Shen, Wenbin Wang, Jun Zhao, Jing Tao, Ulrich Aschauer, Jun Chen, Yimei Zhu, and Jing Zhu. Critical Role of Sc Substitution in Modulating Ferroelectricity in Multiferroic LuFeO3. Nano Lett. 21, 6648−6655(2021).

63. Dae Ho Kim; Ho Nyung Lee; Michael D. Biegalski; Hans M. Christen. Effect of epitaxial strain on ferroelectric polarization in multiferroic BiFeO3 films. Appl. Phys. Lett. 92, 012911 (2008).

64. Fengliang Liu, Changsong Xu, Shoudong Shen, Nana Li, Hangwen Guo, Xujie Lü, Hongjun Xiang, L. Bellaiche, Jun Zhao, Lifeng Yin, Wenge Yang, Wenbin Wang, and Jian Shen. Pressure-induced large enhancement of Néel temperature and electric polarization in the hexagonal multiferroic Lu0.5Sc0.5FeO3. Phy. Rev. B 100, 214408 (2019).

65. Xu. C., Yang. Y., Wang. S. et al. Anomalous properties of hexagonal rare-earth ferrites from first principles. Phy. Rev. B. 89, 205122 (2014).

66. Kishan Sinha, Yubo Zhang, Xuanyuan Jiang, Hongwei Wang, Xiao Wang, Xiaozhe Zhang, Philip J. Ryan, Jong-Woo Kim, John Bowlan, Dmitry A. Yarotski, Yuelin Li, Anthony D. DiChiara, Xuemei Cheng, Xifan Wu, and Xiaoshan Xu. Effects of biaxial strain on the improper multiferroicity in $h-$LuFeO3 films studied using the restrained thermal expansion method. Phys. Rev. B 95, 094110 (2017).

## 2.5 Ferroelectric switching dynamics in h-RFeO3 film

67. Y. Takagi and Yoshihiro Ishibashi, Note on ferroelectric [33] domain switching, J. Phys. Soc. Jpn. 31, 506 (1971).

68. A. Ruff, Z. Y. Li, A. Loidl, J. Schaab, M. Fiebig, A. Cano, Z. W. Yan, E. Bourret, J. Glaum, D. Meier, and S. Krohns, Frequency dependent polarisation switching in h-ErMnO3, Appl. Phys. Lett. 112, 182908 (2018).

69. Shaobo Cheng, Qingping Meng, Myung-Geun Han, Shiqing Deng, Xing Li, Qinghua Zhang, Guotai Tan, Gianluigi A. Botton, and Yimei Zhu. Revealing the Effects of Trace Oxygen Vacancies on Improper Ferroelectric Manganite with In Situ Biasing. Adv. Electron. Mater. 5, 1800827(2019)

70. Myung-Geun Han, Yimei Zhu, Lijun Wu, Toshihiro Aoki, Vyacheslav Volkov, Xueyun Wang, Seung Chul Chae, Yoon Seok Oh, and Sang-Wook Cheong. Ferroelectric Switching Dynamics of Topological Vortex Domains in a Hexagonal Manganite. Adv. Mater. 25, 2415–2421(2013)

71. A. K. Tagantsev, I. Stolichnov, N. Setter, J. S. Cross, and M. Tsukada, Non-Kolmogorov-Avrami switching kinetics [34] in ferroelectric thin films, Phys. Rev. B 66, 214109 (2002).

72. Y. Ishibashi and H. Orihara, A theory of D-E hysteresis loop, Integr. Ferroelectr. 9, 57 (1995).

73. X. Du and I.-W. Chen, Frequency spectra of fatigue of PZT and other ferroelectric thin films, MRS Online Proc. Libr. 493, 311 (1997).

74. Li, X., Yun, Yu et al. Duality of switching mechanisms and transient negative capacitance in improper ferroelectrics. arXiv:2309.14639.

## 2.6 Quantification of K3 distortion from STEM image

75. Megan E. Holtz et al. Topological Defects in Hexagonal Manganites: Inner Structure and Emergent Electrostatics. Nano Lett. 17, 10, 5883–5890(2017).

76. Megan E. Holtz et al. Dimensionality-Induced Change in Topological Order in Multiferroic Oxide Superlattices. Phys. Rev. Lett. 126, 157601 (2022).

77. Mundy, J., Brooks, C., Holtz, M. et al. Atomically engineered ferroic layers yield a room-temperature magnetoelectric multiferroic. Nature 537, 523–527 (2016).

78. Fan, S., Das, H., Rébola, A. *et al.* Site-specific spectroscopic measurement of spin and charge in $(LuFeO_3)_m/(LuFe_2O_4)_1$ multiferroic superlattices. *Nat Commun* **11**, 5582 (2020).

79. Fei-Ting Huang, Xueyun Wang, Sinead M. Griffin, Yu Kumagai, Oliver Gindele, Ming-Wen Chu, Yoichi Horibe, Nicola A. Spaldin, and Sang-Wook Cheong. Duality of Topological Defects in Hexagonal Manganites. Phys. Rev. Lett. 113, 267602(2014).

80. Ren, G., Omprakash, P., Li, X., et al. Polarization Pinning at an Antiphase Boundary in Multiferroic $YbFeO_3$. arXiv:2409.08902

3. Magnetism

81. Lee, S., Pirogov, A., Kang, M. et al. Giant magneto-elastic coupling in multiferroic hexagonal manganites. Nature 451, 805–808 (2008).

82. Tara N. Tošić, Quintin N. Meier, and Nicola A. Spaldin. Influence of the triangular Mn-O breathing mode on magnetic ordering in multiferroic hexagonal manganites. Phys. Rev. Research 4 (2022).

83. Xiang-Bai Chen; Nguyen Thi Minh Hien; D. Lee; S.-Y. Jang; T. W. Noh; In-Sang Yang. Spin exchange interactions in hexagonal manganites $RMnO_3$ (R = Tb, Dy, Ho, Er) epitaxial thin films. Appl. Phys. Lett. 99, 052506 (2011).

3.1 Spin-lattice coupling in h-RFeO3

84. I. V. Solovyev, M. V. Valentyuk, and V. V. Mazurenko. $YMnO_3$ and $LuMnO_3$ from a microscopic point of view. Phys. Rev. B 86, 054407 (2012).

85. Y. S. Tang, S. M. Wang, L. Lin, V. Ovidiu Garlea, Tao Zou, S. H. Zheng, H.-M. Zhang, J. T. Zhou, Z. L. Luo, Z. B. Yan, S. Dong, T. Charlton, and J.-M. Liu. Magnetic structure and multiferroicity of Sc-substituted hexagonal $YbFeO_3$. Phys. Rev. B 103, 174102 (2021)

86. W. Wang, H. Wang, X. Xu, L. Zhu, L. He, E. Wills, X. Cheng, D. J. Keavney, J. Shen, X. Wu, and X. Xu, Crystal field splitting and optical bandgap of hexagonal $LuFeO_3$ films. Applied Physics Letters 101, 241907 (2012).

3.2 K3 distortion and Neel temperature (TN) in h-RFeO3

87. T. Katsufuji, M. Masaki, A. Machida, M. Moritomo, K. Kato, E. Nishibori, M. Takata, M. Sakata, K. Ohoyama, K. Kitazawa, and H. Takagi. Crystal structure and magnetic properties of hexagonal $RMnO_3$($R$=Y, Lu, and Sc) and the effect of doping. Phys. Rev. B 66, 134434 (2002).

88. Junghwan Park, Seongsu Lee, Misun Kang, Kwang-Hyun Jang, Changhee Lee, S. V. Streltsov, V. V. Mazurenko, M. V. Valentyuk, J. E. Medvedeva, T. Kamiyama, and J.-G. Park. Doping dependence of spin-lattice coupling and two-dimensional ordering in multiferroic hexagonal $Y_{1-x}Lu_xMnO_3$ ($0 \leq x \leq 1$). Phys. Rev. B 82, 054428 (2010).

89. M. J. Swamynadhan, Andrew O'Hara, Saurabh Ghosh, and Sokrates T. Pantelides. Engineering Collinear Magnetization in Hexagonal LuFeO3 and Magnetoelectric Control of Skyrmions in Hexagonal 2D Epilayers. Adv. Funct. Mater. 2400195 (2024).

90. Bas B. Van Aken, Jan-Willem G. Bos, Robert A. de Groot, and Thomas T. M. Palstra. Asymmetry of electron and hole doping in YMnO3. Phys. Rev. B 63, 125127 ( 2001).**3.3 K3 distortion and weak ferromagnetism in h-RFeO3**

91. T. Moriya, Anisotropic Superexchange Interaction and Weak Ferromagnetism Phys. Rev. 120, 91 (1960).

92. I. Dzyaloshinsky, A thermodynamic theory of "weak" ferromagnetism of antiferromagnetics. J. Phys. Chem. Solids 4, 241 (1958).

93. F. Keffer, Moriya Interaction and the Problem of the Spin Arrangements in $\beta$MnS Phys. Rev. 126, 896 (1962).**3.4 Effect of structure distortion of FeO5 on single-ion anisotropy**

94. D.-Y. Cho, J.-Y. Kim, B.-G. Park, K.-J. Rho, J.-H. Park, H.-J. Noh, B. J. Kim, S.-J. Oh, H.-M. Park, J.-S. Ahn, H. Ishibashi, S-W. Cheong, J. H. Lee, P. Murugavel, T. W. Noh, A. Tanaka, and T. Jo. Ferroelectricity Driven by Y $d0$-ness with Rehybridization in YMnO3. Phys. Rev. Lett. 98, 217601(2007).

95. Deok-Yong Cho, S.-J. Oh, Dong Geun Kim, A. Tanaka, and J.-H. Park. Investigation of local symmetry effects on the electronic structure of manganites: Hexagonal YMnO3 versus orthorhombic LaMnO3. Phys. Rev. B 79, 035116 (2009)**3.5 The ferrimagnetic order in h-YbFeO3**

96. Cao. S., Sinha. Kishan., Zhang Xin. et al. Electronic structure and direct observation of ferrimagnetism in multiferroic hexagonal YbFeO3 . Phys. Rev. B 95, 224428 (2017).

97. V. N. Antonov and D. A. Kukusta. Electronic structure and x-ray magnetic circular dichroism in the multiferroic oxide $h-$YbFeO3. Phys. Rev. B 99, 104403 (2019).

98. Y. K. Jeong, J. Lee, S. Ahn, S.-W. Song, H. M. Jang, H. Choi, and J. F. Scott, Structurally Tailored Hexagonal Ferroelectricity and Multiferroism in Epitaxial YbFeO3 Thin-Film Heterostructures. J. Am. Chem. Soc. 134, 1450 (2012).

99. A. K. Zvezdin and V. M. Matveev, Sov. Phys. JETP 3, 140 (1972).

100. V. N. Derkachenko, A. M. Kodomtseva, V. A. Timofeeva, and V. A. Khokhlov, JEPT Lett. 20, 104 (1974).

101. N. W. Ashcroft and N. D. Mermin, Solid State Physics (Holt, Rinehart and Winston, New York, 1976).**3.6 Magnetic domains in h-RFeO3**

102. Li, X., Yun, Y., Thind, A.S. et al. Domain-wall magnetoelectric coupling in multiferroic hexagonal YbFeO3 films. Sci Rep 13, 1755 (2023).

4.Magnetoelectric couplings

103.Z. J. Huang, Y. Cao, Y. Y. Sun, Y. Y. Xue, and C. W. Chu. Coupling between the ferroelectric and antiferromagnetic orders in YMnO3. Phys. Rev. B 56 (1997)

104.Hena Das. Coupling between improper ferroelectricity and ferrimagnetism in the hexagonal ferrite LuFeO3. Phys. Rev. Research 5, 013007 ( 2023).

105.M. Ye and D. Vanderbilt, Phys. Rev. B 92, 035107 (2015).

4.1 Magnetoelectric coupling in bulk-state

106. Du, K., Gao, B., Wang, Y. et al. Vortex ferroelectric domains, large-loop weak ferromagnetic domains, and their decoupling in hexagonal (Lu, Sc)FeO3. npj Quant Mater 3, 33 (2018).

107. Gustau Catalan and James F. Scott. Physics and Applications of Bismuth Ferrite. Adv. Mater. 21, 2463–2485 (2009).

108. Heron, J., Bosse, J., He, Q. et al. Deterministic switching of ferromagnetism at room temperature using an electric field. Nature 516, 370–373 (2014).

109. J. T. Heron, M. Trassin,, K. Ashraf, et al. Electric-Field-Induced Magnetization Reversal in a Ferromagnet-Multiferroic Heterostructure. Phys. Rev. Lett. 107, 217202 (2011)

110. Rachel Steinhardt. Ferroic Ferrite Films: An MBE Odyssey. Phd thesis (2020).

4.2 Magnetoelectric coupling in clamped antiferromagnetic domain wall

111. Geng, Y., Lee, N., Choi, Y. J., Cheong, S.-W. & Wu, W. Collective magnetism at multiferroic vortex do-main walls. Nano Lett. 12, 6055–6059 (2012).

112.Kumagai, Y., Spaldin, N. Structural domain walls in polar hexagonal manganites. Nat Commun 4, 1540 (2013).

113.Didrik R. Småbråten, Quintin N. Meier, Sandra H. Skjærvø, Katherine Inzani, Dennis    Meier, and Sverre M. Selbach. Charged domain walls in improper ferroelectric hexagonal manganites and gallates. Phys. Rev. Materials 2, 114405 (2018).

114. K. L. Yang, Y. Zhang, S. H. Zheng, L. Lin, Z. B. Yan, J.-M. Liu, and S.-W. Cheong. Electric field driven evolution of topological domain structure in hexagonal manganites. Phys. Rev. B 96, 144103 ( 2017).

5.Outlook and summary

115. W. Zhong, David Vanderbilt, and K. M. Rabe.First-principles theory of ferroelectric phase transitions for perovskites: The case of BaTiO3. Phys. Rev. B 52, 6301(1995)

116. Cohen, R. Origin of ferroelectricity in perovskite oxides. Nature 358, 136–138 (1992).

117. Jing Li, Andrew H. Comstock, Aeron McConnell, Xin Li, Yu Yun, Dali Sun, and Xiaoshan Xu. Giant interfacial spin Hall angle from Rashba-Edelstein effect revealed by the spin Hall Hanle process. Phys. Rev. B 108, L241403 (2023)

118. Alexander Vogel, Alicia Ruiz Caridad, Johanna Nordlander, Martin F. Sarott, Quintin N. Meier, Rolf Erni, Nicola A. Spaldin, Morgan Trassin, and Marta D. Rossell . Origin of the Critical Thickness in Improper Ferroelectric Thin Films. ACS Appl. Mater. Interfaces 15, 14, 18482–18492 (2023)

119. Akbashev, A., Roddatis, V., Vasiliev, A. et al. Reconstruction of the polar interface between hexagonal $LuFeO_3$ and intergrown $Fe_3O_4$ nanolayers. Sci Rep 2, 672 (2012).

120. Daesu Lee, A Yoon, SY Jang, J-G Yoon, J-S Chung, M Kim, JF Scott, TW Noh. Giant flexoelectric effect in ferroelectric epitaxial thin films. Phys. Rev. Lett. 107, 057602 (2011)

121. Zubko, P., Catalan, G. & Tagantsev, A. K. Flexoelectric Effect in Solids. Annu Rev Mater Res 43, 387–421 (2013).

122. Xin Li, Guodong Ren, Yu Yun, Arashdeep Singh Thind, Amit Kumar Shah, Abbey Bowers, Rohan Mishra, Xiaoshan Xu. Improper flexoelectricity in hexagonal rare-earth ferrites. arXiv:2409.17022

123. S. M. Griffin, M. Lilienblum, K. T. Delaney, Y. Kumagai, M. Fiebig, and N. A. Spaldin. Scaling Behavior and Beyond Equilibrium in the Hexagonal Manganites. Phys. Rev. X **2**, 041022 (2012)

124. Lin, S.-Z. et al. Topological defects as relics of emergent continuous symmetry and Higgs condensation of disorder in ferroelectrics. Nat. Phys. 10, 970–977 (2014).

125. Xueyun Wang, Maxim Mostovoy, Myung-Geun Han, Yoichi Horibe, T Aoki, Yimei Zhu, S-W Cheong. Unfolding of vortices into topological stripes in a multiferroic material. Phys. Rev. Lett. 112, 247601(2014).

126. Shaobo Cheng; Dong Zhang; Shiqing Deng; Xing Li; Jun Li; Guotai Tan, Yimei Zhu; Jing Zhu. Domain configurations in dislocations embedded hexagonal manganite systems: From the view of graph theory. Appl. Phys. Lett. 112, 162905 (2018)

127. S. C. Chae, Y. Horibe, D. Y. Jeong, S. Rodan, N. Lee, and S.-W. Cheong, Self-organization, condensation, and annihilation of topological vortices and antivortices in a multiferroic. Proc. Natl. Acad. Sci. 107, 21366 (2010).

128. Hyo Ju Park, Roland Yingjie Tay, Xiao Wang, Wen Zhao, Jung Hwa Kim, Rodney S. Ruoff, Feng Ding, Edwin Hang Tong Teo, and Zonghoon Lee. Double-Spiral Hexagonal Boron Nitride and Shear Strained Coalescence Boundary. Nano Lett. 19, 7, 4229–4236 (2019).

129. Yuzhou Zhao, Chenyu Zhang, Daniel D Kohler, Jason M Scheeler, John C Wright, Paul M Voyles, Song Jin. Supertwisted spirals of layered materials enabled by growth on non-Euclidean surfaces. Science 370, 6515 442-445 (2020)

130. Wang, ZJ., Kong, X., Huang, Y. et al. Conversion of chirality to twisting via sequential one-dimensional and two-dimensional growth of graphene spirals. Nat. Mater. 23, 331–338 (2024).

131. Han, L., Addiego, C., Prokhorenko, S. et al. High-density switchable skyrmion-like polar nanodomains integrated on silicon. Nature 603, 63–67 (2022).

132. Sánchez-Santolino, G., Rouco, V., Puebla, S. et al. A 2D ferroelectric vortex pattern in twisted BaTiO3 freestanding layers. Nature 626, 529–534 (2024).

133. : Robert Streubel et al. Magnetism in curved geometries J. Phys. D: Appl. Phys. 49 363001(2016).

134. Lu, D., Baek, D., Hong, S. et al. Synthesis of freestanding single-crystal perovskite films and heterostructures by etching of sacrificial water-soluble layers. Nature Mater 15, 1255–1260 (2016).

135. Jinfeng Zhang, Ting Lin, Ao Wang, Xiaochao Wang, Qingyu He, Huan Ye, Jingdi Lu, Qing Wang, Zhengguo Liang, Feng Jin, Shengru Chen, Minghui Fan, Er-Jia Guo, Qinghua Zhang, Lin Gu, Zhenlin Luo, Liang Si, Wenbin Wu, Lingfei Wang. Super-tetragonal $Sr_4Al_2O_7$ as a sacrificial layer for high-integrity freestanding oxide membranes. Science 383 (6681), 388-394 (2024)

136. Ji, D., Cai, S., Paudel, T.R. et al. Freestanding crystalline oxide perovskites down to the monolayer limit. Nature 570, 87–90 (2019).

137. Xu, R., Huang, J., Barnard, E.S. et al. Strain-induced room-temperature ferroelectricity in $SrTiO_3$ membranes. Nat Commun 11, 3141 (2020).

138. Haoying Sun, Jiahui Gu, Yongqiang Li, Tula R. Paudel, Di Liu, Jierong Wang, Yipeng Zang, Chengyi Gu, Jiangfeng Yang, Wenjie Sun, Zhengbin Gu, Evgeny Y. Tsymbal, Junming Liu, Houbing Huang, Di Wu, and Yuefeng Nie."Prominent size effects without a depolarization field observed in ultrathin ferroelectric oxide membranes." Phys. Rev. Lett. 130,126801(2023)

139. Reji Thomas et al Multiferroic thin-film integration onto semiconductor devices. J. Phys.: Condens. Matter 22 423201 (2010)

140. T. Mikolajick, U. Schroeder , and S. Slesazeck. The Past, the Present, and the Future of Ferroelectric Memories. IEEE TRANSACTIONS ON ELECTRON DEVICES 67, 4 (2020)

141. Ik-Jyae Kim, Jang-Sik Lee. Ferroelectric Transistors for Memory and Neuromorphic Device Applications. Adv. Mater. 35, 2206864 (2023)

142. Meier, D., Selbach, S.M. Ferroelectric domain walls for nanotechnology. Nat Rev Mater 7, 157–173 (2022).

143. Hyeon Han, Donghoon Kim, Kanghyun Chu et al. Enhanced Switchable Ferroelectric Photovoltaic Effects in Hexagonal Ferrite Thin Films via Strain Engineering. ACS Appl. Mater. Interfaces 10, 1846−1853(2018)